\documentstyle[epsfig]{cupconf}

\ifoldfss
\else
  \ifnfssone
    \newmathalphabet{\mathit}
      \addtoversion{normal}{\mathit}{cmr}{m}{it}
      \addtoversion{bold}{\mathit}{cmr}{bx}{it}
    \newmathalphabet{\mathcal}
      \addtoversion{normal}{\mathcal}{cmsy}{m}{n}
    \else
    \ifnfsstwo
    \fi
  \fi
\fi

\def\gtrsim{\mathrel{\hbox{\rlap{\hbox{\lower4pt\hbox{$\sim$}}}\hbox{$>$}}}}
\def\lesssim{\mathrel{\hbox{\rlap{\hbox{\lower4pt\hbox{$\sim$}}}\hbox{$<$}}}}

\def\hexnumber#1{\ifcase#1 0\or1\or2\or3\or4\or5\or6\or7\or8\or9\or
 A\or B\or C\or D\or E\or F\fi }

\makeatletter
\ifx\CUP@mtlplain@loaded\undefined
\else
\fi
\makeatother
 \makeatletter
 \ifx\CUP@mtlplain@loaded\undefined
   \font\tenbmi=cmmib10 at 10pt
   \font\sevenbmi=cmmib10 at 7pt
   \font\fivebmi=cmmib10 at 5pt

   \newfam\bmifam
   \textfont\bmifam=\tenbmi
   \scriptfont\bmifam=\sevenbmi
   \scriptscriptfont\bmifam=\fivebmi
   
 \fi
 \makeatother
\ifnfsstwo

\fi
\ifnfssone

\fi
\ifoldfss

\fi

\mathchardef\varLambda="0103

\makeatletter
\ifx\CUP@mtlplain@loaded\undefined
\else
\fi
\makeatother
\makeatletter
\ifx\CUP@mtlplain@loaded\undefined
  \font\tenbms=cmbsy10
  \font\sevenbms=cmbsy10 at 7pt
  \font\fivebms=cmbsy10 at 5pt
  \newfam\bmsfam
  \textfont\bmsfam=\tenbms
  \scriptfont\bmsfam=\sevenbms
  \scriptscriptfont\bmsfam=\fivebms

  \edef\bsy@{\hexnumber\bmsfam}
  \mathchardef\bnabla="0\bsy@72
\fi
\makeatother
\def\ltsima{$\; \buildrel < \over \sim \;$}
\def\simlt{\lower.5ex\hbox{\ltsima}}
\def\gtsima{$\; \buildrel > \over \sim \;$}
\def\simgt{\lower.5ex\hbox{\gtsima}}
\def\lya{Ly$\alpha$}
\def\kms{\,km~s$^{-1}$}

\newcommand{\Hb}{\mbox{H$\beta$}}

\begin{document}
\title[Element Abundances through the Cosmic Ages]
{Element Abundances through the Cosmic Ages\footnote{{\tt Lectures given at the XIII Canary Islands
Winter School of Astrophysics {\it `Cosmochemistry: The
Melting Pot of Elements'}. Available in electronic form from
http://www.ast.cam.ac.uk/\~\,pettini/canaries13}}
}

\author[Max Pettini]%
{M\ls A\ls X\ns P\ls E\ls T\ls T\ls
 I\ls N\ls I$^1$}

\affiliation{$^1$
Institute of Astronomy, University of Cambridge\\
Madingley Road, Cambridge, UK}

\setcounter{page}{1}


\ifnfssone
\else
  \ifnfsstwo
  \else
    \ifoldfss
      \let\mathcal\cal
      \let\mathrm\rm
      \let\mathsf\sf
    \fi
  \fi
\fi

\maketitle

\begin{abstract}
The horizon for studies of element abundances has 
expanded dramatically in the last ten years.
Once the domain of astronomers concerned chiefly with stars and
nearby galaxies, this field has now become a key component
of observational cosmology, as technological advances
have made it possible to measure the
abundances of several chemical elements in a variety of environments
at redshifts up to $z \simeq 4$, when the universe
was in its infancy. In this series of 
lectures I summarise current knowledge on the chemical
make-up of distant galaxies observed directly in
their starlight, and of interstellar and intergalactic gas
seen in absorption against the spectra of bright background sources.
The picture which is emerging is one where the
universe at $z = 3$ already included many of the constituents
of today's galaxies---even at these early times we see
evidence for Population I and II stars, while the `smoking gun'
for Population III objects may be hidden in the chemical
composition of the lowest density regions of the intergalactic medium, 
yet to be deciphered. 
\end{abstract}

\section{Introduction}

One of the exciting developments in observational cosmology
over the last few years has been the ability to extend studies
of element abundances from the local universe to high redshifts.
Thanks largely to the new opportunities offered by the
Keck telescopes, the Very
Large Telescope facility at the European Southern Observatory, 
and most recently the Subaru telescope,
we find ourselves in the 
exciting position of being
able, for the first time, to detect and measure a wide 
range of chemical elements directly in stars, H~II regions, cool
interstellar gas and hot intergalactic medium, all observed when
the universe was only $\sim 1/15$ of its present age. Our
simple-minded hope is that, by moving back to a time when the
universe was young, clues to the nature, location, and epoch of
the first generations of stars may be easier to interpret than in
the relics left today, some 12~Gyrs later. Furthermore,
the metallicities of different structures in the
universe and their evolution with redshift are key factors 
to be considered in our attempts to track
the progress of galaxy formation through the cosmic
ages. 

In the last few years, work on chemical abundances at high
redshifts has concentrated on four main components of the
young universe: Active Galactic Nuclei (AGN), that is 
quasars (QSOs) and Seyfert galaxies; 
two classes of QSO absorption lines,
the damped \lya\ systems (DLAs) and the \lya\
forest; and on galaxies detected directly via
their starlight, also referred to as Lyman break galaxies. 
In these series of lectures I will review results pertaining to 
the last three; for abundance determinations in the emission line
regions of AGN and associated absorbers I refer the
interested reader to the excellent recent overview by 
Hamann \& Ferland (1999).

\subsection{Some Basic Concepts}

For many years our knowledge of the distant universe
relied almost exclusively on QSO absorption lines. 
It is only relatively recently that we
have learnt to identify directly `normal' galaxies;
up until 1995 the only objects known at high $z$ were
QSOs and powerful radio galaxies. The technique
of QSO absorption line spectroscopy, illustrated in
Figure 1, is potentially very powerful. As we shall 
see, it allows accurate measurements
of many physical properties of the interstellar medium 
(ISM) in galaxies and the intergalactic 
medium (IGM) between galaxies. 
The challenge, however, is to relate this
wealth of data, which refer to gas along a very narrow
sightline, to the global properties of the absorbers.
In a sense, all the information we obtain from QSO
absorption line spectroscopy is of an {\it indirect} nature;
if we could detect the galaxies themselves,
our inferences would be on a stronger empirical basis.
%
\begin{figure} 
\vspace*{0.75cm} 
\hspace*{0.55cm}
\psfig{figure=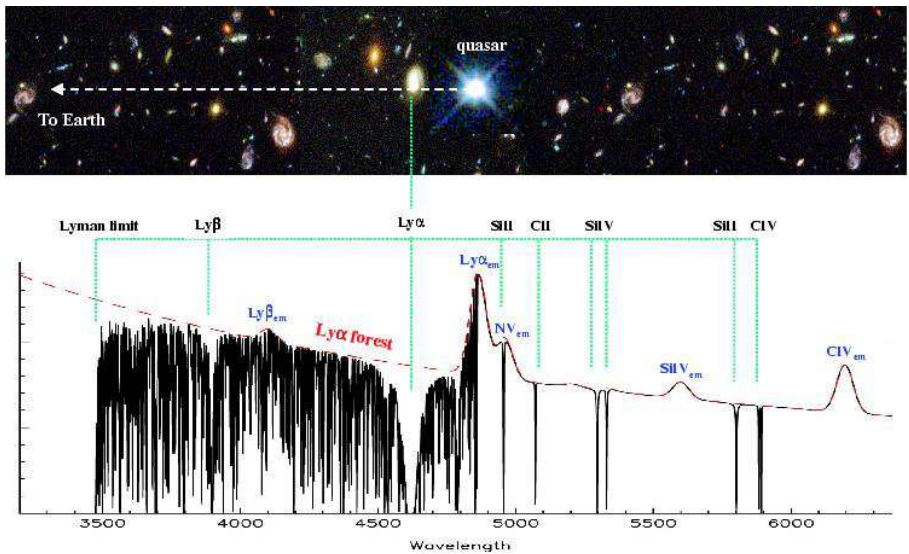,width=12.0cm} 
\vspace*{0.5cm} 
\caption{
The technique of QSO absorption line spectroscopy
is illustrated in this montage (courtesy of John Webb).
QSOs are among the brightest and most distant objects
known. On the long journey from its source to our
telescopes on Earth, the light from a background QSO
intercepts galaxies and intergalactic matter which
happen to lie along the line of sight (and are therefore
at lower absorption redshifts, $z_{\rm abs}$, than the 
QSO emission redshift, $z_{\rm em}$. Gas in these
structures leaves a clear imprint in the 
spectrum of the QSO in the form of narrow absorption
lines. The task of astronomers working in this field
has been to relate the characteristics of the absorption
lines to the properties of the intervening
galaxies which are normally
too faint to be detected directly.
}
\end{figure}

In deriving chemical abundances in QSO absorbers and 
high redshift galaxies, we shall make use of some of the same 
techniques which are applied locally to interpret the spectra
of stars, cool interstellar gas and H~II regions. These 
methods are discussed extensively in other articles
in this volume, particularly those by  Don Garnett, David Lambert,
and Grazyna Stasinska, and will therefore not be 
repeated here. 
The derivation of ion column densities from the profiles and
equivalent widths of interstellar absorption lines is discussed 
in a number of standard textbooks, as well as a recent volume 
in this series (Bechtold 2002).

When measuring element abundances in different astrophysical
environments, we shall often compare them to the 
composition of the solar system 
determined either from photospheric lines in the solar spectrum
or, preferably, from meteorites. 
The standard solar abundance
scale continues to be refined; here we adopt the 
compilation  by Grevesse \& Sauval (1998) with the recent
updates by Holweger (2001). We use the standard notation
[X/Y] = $\log$(X/Y)$_{\rm obs} - \log$(X/Y)$_{\odot}$
where (X/Y)$_{\rm obs}$ 
denotes the abundance of element X relative to element Y
in the system under observation---be it stars, interstellar
gas or the intergalactic medium---and 
(X/Y)$_{\odot}$ is their relative abundance in the solar system.

Furthermore, it is important to remember that when element
abundances are measured in the interstellar gas of the Milky Way,
it is usually found that [X/H]\,$ < 0$. This deficiency
is not believed to be intrinsic, but rather
reflects the proportion of heavy elements that has condensed
out of the gas phase to form dust grains (and therefore
no longer absorbs starlight via discrete atomic transitions).
As discussed in the review by Savage \& Sembach
(1996), the `missing' fraction varies from element
to element, reflecting the ease with which different constituents
of interstellar dust are either incorporated into the grains
or released from them. In particular, O, N, S, and Zn show little
affinity for dust and are often present in the gas in near-solar
proportions; Si, Fe and most Fe-peak elements, on the other hand,
can be depleted by large and varying amounts depending on the 
physical conditions---past and present---of the interstellar clouds
under study.

Unless otherwise stated, we shall use today's 
`consensus' cosmology (e.g. Turner 2002) with 
$H_0 = 65 $\,km~s$^{-1}$~Mpc$^{-1}$ (and hence $h = 0.65$), 
$\Omega_{\rm baryons} = 0.022\,h^{-2}$, 
$\Omega_{\rm M} = 0.3$, and $\Omega_{\Lambda} = 0.7$.
Table 1 shows the run of look-back time with redshift for
this cosmology. Note that with the above cosmological parameters
the age of the universe is 14.5\,Gyr, consistent
with recent estimates of the ages of globular
clusters (Krauss \& Chaboyer 2001). 
When in these lecture notes we refer to `high' redshifts 
we usually mean $z = 3-4$ which correspond to look-back times
of 12--13\,Gyr, when the universe had only 15--10\% of its
present age. Epochs when $z$ was $ \leq 1$ are generally referred to as
`intermediate' or `low' redshifts, even though they correspond
to look-back times of up to about 60\% of the current age
of the universe.

%
%
\begin{center}
\vspace{0.75cm}
\begin{tabular}{c c c}
\multicolumn{3}{c}{{\bf Table 1.} Lookback time vs. redshift in the adopted
cosmology
}\\
\hline
Redshift  & Lookback Time (Gyr) & Lookback time ($t/t_{\infty}$) \\
\hline
0       & 0    & 0 \\
0.5     & 5.4  & 0.37 \\
1       & 8.3  & 0.57 \\
2       & 11.0 & 0.76 \\
3       & 12.2 & 0.84 \\
4       & 12.9 & 0.89 \\
5       & 13.3 & 0.92 \\
6       & 13.5 & 0.93 \\
10      & 14.0 & 0.97 \\
$\infty$   & 14.5 & 1.00\\
\hline
\end{tabular}
\end{center}
\vspace{0.5cm}

\section{Damped \lya\ Systems}

\subsection{What Are They?}

I am often asked this question, and the truthful answer
is: ``We do not know''. Spectroscopically, DLAs are 
straightforward to identify (see Figure 2). The large
equivalent widths and characteristic damping wings
which signal column densities of absorbing neutral hydrogen
in excess of $N$(H~I)\,$= 2 \times 10^{20}$\,cm$^{-2}$
are easy to recognise even in spectra of moderate 
resolution and signal-to-noise ratio (significantly 
worse than those of the spectrum reproduced in Figure 2).

\begin{figure} 
\vspace*{-1.5cm} 
\hspace*{-1.5cm}
\psfig{figure=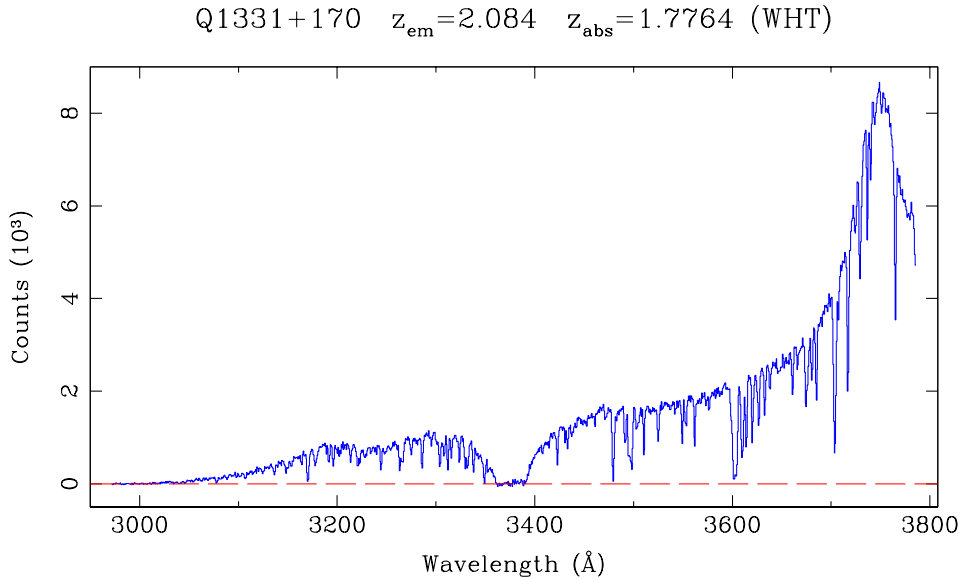,width=12.0cm,angle=270} 
\vspace*{-2.0cm} 
\caption{The strong absorption feature centred near 3375\,\AA\ 
in the near-ultraviolet (UV) spectrum of the bright QSO
Q1331+170 is a 
good example of a damped \lya\ line, in this case produced
by a column density of neutral hydrogen atoms 
$N$(H~I)\,$= 1.5 \times 10^{21}$\,cm$^{-2}$.
This spectrum was recorded in the early 1990s with the 
Image Photon Counting System on the ISIS spectrograph
of the 4.2\,m William Herschel telescope on La Palma
(Pettini et al. 1994).}
\end{figure}

The galaxies producing DLAs, however, have proved difficult
to pin down. Wolfe and collaborators, who were the first
to recognise DLAs as a class of QSO absorbers of special
significance for the study of the high redshift universe
(e.g. Wolfe et al. 1986), proposed from the outset that
they are the progenitors of present-day spiral galaxies,
observed at a time when most of their baryonic mass was 
still in gaseous form. The evidence supporting this 
scenario, however, is mostly indirect. For example,
Prochaska \& Wolfe (1998) showed that the profiles of
the metal absorption lines in DLAs are consistent with
the kinematics expected from large, rotating, thick disks,
but others have claimed that this interpretation
is not unique (Haehnelt, Steinmetz, \& Rauch 1998; 
Ledoux et al. 1998).

Imaging studies at high redshift are only now beginning
to identify some of the absorbers (Prochaska et al. 2002;
M\o ller et al. 2002). At $z < 1$, where the imaging is 
easier (an example is reproduced in Figure 3), 
it appears that DLA galaxies are a very `mixed bag',
which includes a relatively high proportion of low surface
brightness and low luminosity galaxies (some so faint
that they remain undetected in their stellar 
populations, e.g. Steidel et al. 1997; Bouch\'{e} et al.
2001), as well as more `normal' 
spirals (Boissier, P\'{e}roux, \& Pettini 2002).

%
\begin{figure}
\vspace*{0.55cm}
\hspace*{3.30cm}
\includegraphics[width=0.5\textwidth]{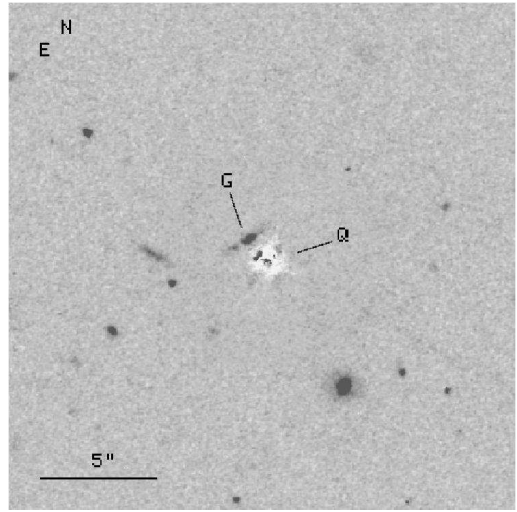}
\vspace*{0.5cm}
\caption[]{(Reproduced from Pettini et al. 2000a).
WFPC2 F702W exposure of the field of Q0058+019 (PHL~938).
A model point spread function has
been subtracted from the QSO image (labelled `Q'),
revealing the presence of a galaxy (labelled `G')
approximately 1.2 arcseconds to the NE
of the QSO position.
Given its proximity, this is likely to be the damped
Ly\,$\alpha$ absorber at $z = 0.61251$.
Residual excess absorption of a diffraction spike cuts across the
galaxy image. When this processing artifact is taken
into account, the candidate absorber appears to be a low
luminosity ($L \simeq 1/6\,L^*$) late-type galaxy seen at high
inclination, $i \approx 65^{\circ}$, at a projected
separation of $6\,h^{-1}$\,kpc from the QSO sightline.
}
\end{figure}

Much was made in the early 1990s of the apparent
correspondence between the neutral gas mass traced by DLAs 
at high redshift 
($\Omega_{\rm DLA}\, h \simeq 1.2 \times 10^{-3}$,
when expressed as a fraction of the critical density)
and today's luminous stellar mass, leading to suggestions
that the former are the material out of which the
latter formed (e.g. Lanzetta 1993). 
However, the apparent decrease in 
$\Omega_{\rm DLA}$ from $z = 3$ to 0, upon which 
this picture was based, has not been confirmed
with more recent and more extensive samples
(Pettini 2001). As can be seen from Figure 4,
current data are consistent with
an approximately constant value $\Omega_{\rm DLA}$ 
over most of the Hubble time, and this includes 
the most recent estimates of $\Omega_{\rm H~I}$ 
in the local universe from 21\,cm surveys
(Rosenberg \& Schneider 2002; not shown in Figure 4).
Perhaps DLAs pick out a particular stage in the evolution
of galaxies, when their dimensions in high surface
density of neutral gas are largest, and 
it may be the case that different
populations of galaxies pass through this stage at
different cosmic epochs.

%
%
\begin{figure} 
\vspace*{-1.75cm} 
\hspace*{-0.25cm}
\epsfig{figure=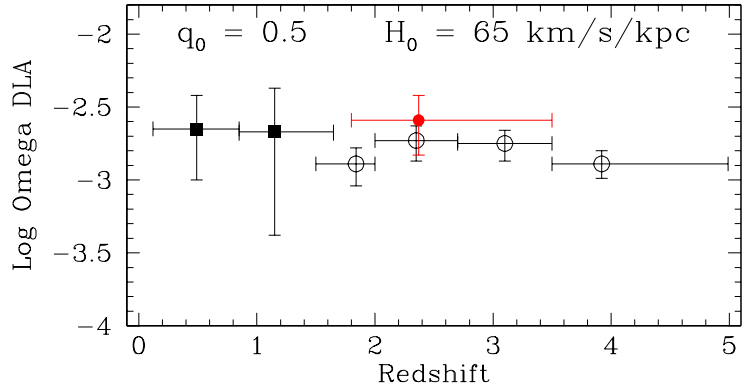,width=10cm,angle=270}
\vspace*{-2.0cm} 
\caption[]{
Recent estimates of the 
mass density of neutral gas traced by damped \lya\ systems,
expressed as a fraction of the critical density. 
The filled squares are from Rao \& Turnshek (2000);
the open circles from Storrie-Lombardi \& Wolfe (2000)
and from P\'{e}roux et al. (2002); while the filled
circle is from the CORALS survey by Ellison et al. (2001).
}
\end{figure}

\subsection{Why Do We Care?}

While we would obviously like to know more clearly
which population(s) of galaxies DLAs are associated with,
this issue does not detract from the importance
of this class of QSO absorbers for studies of
element abundances, for the following reasons.

First, with neutral hydrogen column densities
$N$(H~I)$\geq 2 \times 10^{20}$~cm$^{-2}$,
DLAs are the `heavy weights'
among QSO absorption systems,
at the upper end of the distribution
of values of $N$(H~I) 
which spans 10 orders of magnitude 
for all absorbers (see Figure 5).
Over this entire range, 
$f(N_{\rm H~I})$---defined as the number of 
absorbing systems per unit redshift path 
per unit column density---can be fitted with a single
power law of the form
\begin{equation}
f(N_{\rm H~I}) = B \times {N_{\rm H~I}^{-\beta}}
\end{equation}
with $\beta \simeq 1.5$ 
(Tytler 1987; Storrie-Lombardi \& Wolfe 2000).
While the most numerous absorbers are those
with the lowest column densities
(a turn-over at low values of $N$(H~I)
has yet to be found), the high column densities
of DLAs more than compensate for their relative paucity.
More specifically, so long as $\beta < 2$,
the integral of the column density distribution
\begin{equation}
\Omega_{\rm H~I} = \frac{H_0}{c} \,  \frac{\mu \,  m_{\rm H}}{\rho_{\rm crit}} \, \int_{N_{\rm min}}^{N_{\rm max}} N\, f(N)\, dN
= \frac{H_0}{c} \,  \frac{\mu \,  m_{\rm H}}{\rho_{\rm crit}} \, \frac{B}{2- \beta} \, \left( N_{\rm max}^{2- \beta} - N_{\rm min}^{2- \beta} \right )
\end{equation}
is dominated by $N_{\rm max}$, i.e. by DLAs (Lanzetta 1993).
In eq. (2.2), $H_0$ is the Hubble constant, $c$ is the speed of light, 
$m_{\rm H}$ is the mass of the hydrogen atom,
$\mu$ is the mean atomic weight per baryon 
($\mu = 1.4$ for solar abundances; Grevesse \& Sauval 1998)
and $\rho_{\rm crit}$ is the closure density
\begin{equation}
\rho_{\rm crit} = \frac{3\, H_0^2}{8 \, \pi \, G} = 1.96 \times 10^{-29} h^2\,{\rm g}~{\rm cm}^{-3}
\end{equation}
where $h$ is the Hubble constant in units of 
100\,km~s$^{-1}$~Mpc$^{-1}$.

%
%
\begin{figure}
\vspace*{-0.75cm}
\centerline{\psfig{figure=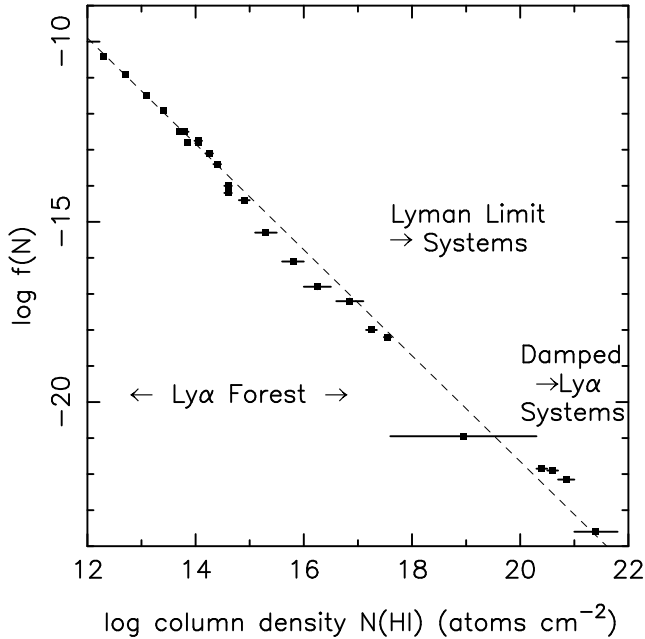,width=10cm,angle=0}}
\caption{(Reproduced from Storrie-Lombardi \& Wolfe 2000).
The column density distribution function of neutral
hydrogen for all QSO absorbers spans ten orders of magnitude,
from log\,$N$(H~I) = 12 to 22 and can be adequately described
by a single power law  $f(N_{\rm H~I}) \propto N_{\rm H~I}^{-1.5}$,
shown by the dashed line.}
\end{figure}

As a consequence, the mean metallicity of DLAs is
the closest measure we have of the global degree of metal enrichment 
of neutral gas in the universe at a given epoch, 
{\it irrespectively of the precise nature of the absorbers},
a point often emphasised by Mike Fall and his collaborators
(e.g. Fall 1996). Of course this only applies if there are
no biases which exclude any
particular type of high redshift object 
from our H~I census.

Second, it is possible to determine
the abundances of a wide variety of elements 
in DLAs with higher precision than in most
other astrophysical environments in the distant
universe.
In particular, echelle spectra obtained with
large telescopes can yield abundance measures
accurate to 10--20\% (e.g. Prochaska \& Wolfe 2002),
because: (a) the damping wings of the \lya\
line are very sensitive to the column density of H~I;
(b) several atomic transitions are often available for
elements of interest; and (c) ionisation corrections are
normally small, because the gas is mostly neutral 
and the major ionisation stages are observed directly
(Vladilo et al. 2001).
Dust depletions can be a complication, but even these
are not as severe in DLAs as in the local interstellar
medium (Pettini et al. 1997a) and can be accounted for
with careful analyses (e.g. Vladilo 2002a).
Thus, abundance studies in DLAs complement in a very
effective way the information provided locally by
stellar and nebular spectroscopy and, as we shall see,
can offer fresh clues to the nucleosynthesis of 
elements,
particularly in metal-poor environments which are
difficult to probe in the nearby universe.
DLAs are also playing a role in the determination
of the primordial abundances of the light elements,
as discussed by Gary Steigman in this volume
(see also Tytler et al. 2000 and Pettini \& Bowen 2001).

\subsection{The Metallicity of DLAs}

Even before the advent of 8-10\,m class telescopes,
it was realised that the metal and dust content of DLAs
could be investigated effectively by targeting 
a pair of (fortuitously) closely spaced 
multiplets, Zn~II~$\lambda\lambda 2025,2062$ and
Cr~II~$\lambda\lambda 2056,2062,2066$ (Meyer, Welty, \& York 1989;
Pettini, Boksenberg, \& Hunstead 1990).
The key points here are that while Zn is essentially undepleted
in local diffuse interstellar clouds, Cr is mostly locked
up in dust grains (Savage \& Sembach 1996). Consequently, 
the ratio $N$(Zn~II)/$N$(H~I) observed in DLAs, when
compared with the solar abundance of Zn, yields
a direct measure of the degree of metal enrichment
(in H~I regions Zn is predominantly singly ionised, 
and the ratios Zn~I/Zn~II and Zn~III/Zn~II are both $\ll 1$).
On the other hand, a deficit---if one is found---of 
$N$(Cr~II)/$N$(Zn~II) compared to the solar relative abundances
of these two elements would measure the
extent to which refractory elements have condensed
into solid form in the interstellar media traced by DLAs 
and, by inference, be an indication of the presence of 
dust in these early galaxies (Cr is also singly
ionised in H~I gas).

%
%
\begin{figure} 
\vspace*{-4.0cm} 
\hspace*{-4.0cm}
\epsfig{figure=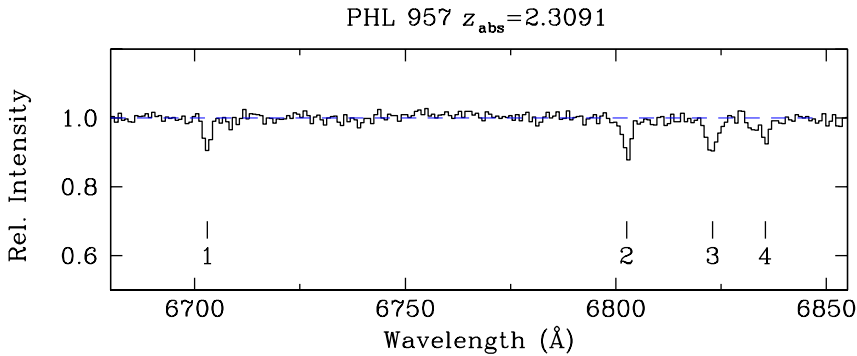,width=15cm,angle=270}
\vspace*{-5.25cm} 
\caption[]{
Portion of the Palomar spectrum of the bright QSO PHL~957
recorded by Pettini et al. (1990), encompassing the region 
of the Zn~II and Cr~II absorption lines in the $z_{\rm abs} = 2.3091$ DLA.
The vertical tick marks indicate the positions of the lines as follows.
Line 1:~Zn~II~$\lambda 2025.483$; 
line 2:~Cr~II~$\lambda 2055.596$; 
line 3:~Cr~II~$\lambda   2061.575$~+~Zn~II~$\lambda  2062.005$  (blended);  
and  line 4:~Cr~II~$\lambda 2065.501$.  
The spectrum has been normalised to the
underlying QSO continuum and is 
shown on an expanded vertical scale.
}
\end{figure}

From the point of view of chemical evolution, both Zn and Cr
closely trace Fe in Galactic stars with metallicities
[Fe/H] between 0 (i.e. solar) and $-2$ (1/100 of solar; 
Sneden, Gratton, \& Crocker 1991, McWilliam et al. 1995).
Additional advantages are the convenient rest wavelengths of the Zn~II
and Cr~II transitions, which at $z = 2 - 3$ (where
DLAs are most numerous in current samples) are redshifted into a 
easily observed portion of the optical spectrum,
and the inherently weak nature of these lines which ensures
that they are nearly always on the linear part of the 
curve of growth, where 
column densities can be derived with confidence  
from the measured equivalent widths (e.g. Bechtold 2002).

All in all, a small portion of the red spectrum of a QSO
with a damped \lya\ system has the potential of providing 
some important chemical clues on the nature of these absorbers
and the evolutionary status of the population of galaxies
they trace. One of the first detections of Zn and Cr in a DLA
is reproduced in Figure 6. However, data of sufficiently high
signal-to-noise ratio to detect the weak absorption
lines of interest (or to place interesting upper limits
on their equivalent widths) typically require nearly
one night of observation per QSO with a 4\,m class telescope.
Consequently, it was necessary to wait until the mid-1990s before
a sufficiently large sample of Zn and Cr measurements in DLAs 
could be assembled.

%
%
\begin{figure} 
\vspace*{-4.0cm} 
\hspace*{-2.75cm}
\epsfig{figure=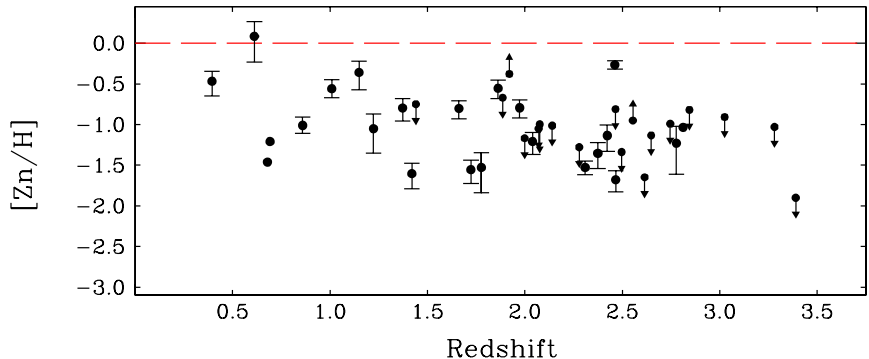,width=14cm,angle=270}
\vspace*{-4.25cm} 
\caption[]{Plot of the abundance of Zn against redshift for the
full sample of 41 DLAs from the surveys by Pettini and collaborators.
Abundances are measured on a log scale relative to the
solar value shown by the broken line at [Zn/H] = 0.0;
thus a point at [Zn/H] = $-1.0$ corresponds to 
a metallicity of 1/10 of solar. Upper
limits, corresponding to non-detections of the  Zn~II lines,
are indicated by downward-pointing arrows.
Upward-pointing arrows denote lower limits 
in two cases where the Zn~II lines are sufficiently
strong that saturation may be important.
}
\end{figure}

Figure 7 shows the current data set from the survey by our group
(e.g. Pettini et al. 1997b; 1999). There are several interesting
conclusions which can be drawn from these results.
\begin{enumerate}
\item Damped \lya\ systems are generally metal poor at all
redshifts sampled. Evidently DLAs arise in galaxies
at early stages of chemical evolution, 
since nearly all of the points in Figure 7 lie below
the line of solar abundance.
The statistic of relevance here is the 
the column density-weighted mean abundance of Zn 
\begin{equation}
	{\rm [} \langle{\rm Zn/H}_{\rm DLA}\rangle {\rm ]} = 
        {\rm log}\langle{\rm(Zn/H)}_{\rm DLA}\rangle-{\rm log~(Zn/H)}_{\odot}, 
	\label{}
\end{equation}
where
\begin{equation}
	\langle{\rm (Zn/H)}_{\rm DLA}\rangle = 
        \frac{\sum\limits_{i=1}^{n} N{\rm(Zn~II)}_i}
        {\sum\limits_{i=1}^{n} N{\rm(H~I)}_i} , 
	\label{}
\end{equation}
and the summations in eq. (2.5) are over 
the $n$ DLA systems in a given sample.
In this way, by counting all the Zn atoms 
per unit cross-section (cm$^{-2}$) 
and dividing by the total column density of neutral hydrogen
we find that DLAs have a typical metallicity
of only $\sim 1/13$ of solar 
(${\rm [} \langle{\rm Zn/H}_{\rm DLA}\rangle {\rm ]} = -1.13$).

\item There appears to be a large range in the values of metallicity 
reached by different galaxies at the same redshift.  
Values of [Zn/H] in Figure 7 span nearly two orders
of magnitude, pointing to 
a protracted `epoch of galaxy formation' and to the fact that
chemical enrichment probably proceeded at different rates in different 
DLAs. The wide dispersion in metallicity goes hand
in hand with the diverse morphology of the DLA galaxies
which have been imaged at $z < 1$, as discussed earlier (\S2.1).

When the metallicity distribution of damped \lya\ systems
is compared with those of different stellar populations of the Milky Way, 
we find that is broader and peaks at 
lower metallicities than those of either thin or thick disk
stars (Figure 8).  At the time when our Galaxy's metal enrichment was at levels 
typical of DLAs, its kinematics were closer to those of the 
halo and bulge than a rotationally supported disk.
This finding is at odds with the 
proposal that most DLAs are large disks with rotation velocities in excess of 
200~\kms, put forward by Prochaska \& Wolfe (1998).

%
\begin{figure}
\centering
\vspace*{-0.5cm}
\hspace*{0.25cm}
\includegraphics[width=0.75\textwidth]{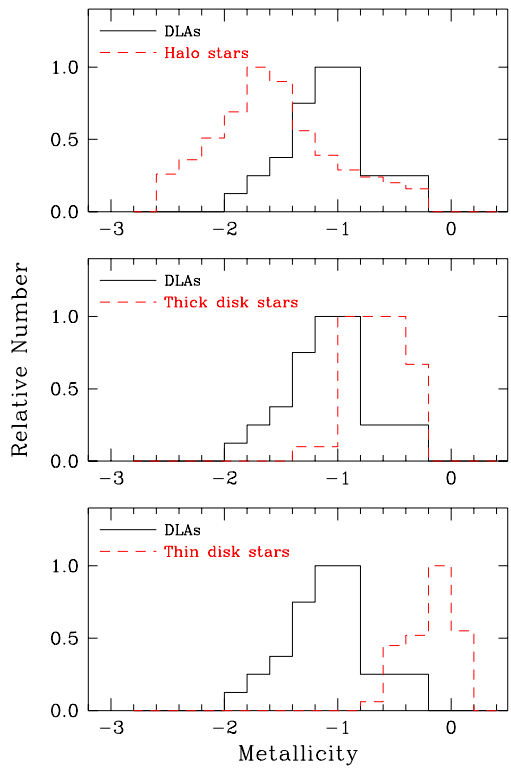}
\vspace*{-1.2cm}
\caption[]{Metallicity distributions, normalised to unity,
of DLAs at $z \simeq 2 - 3$ 
and of stars belonging to the disk 
(Wyse \& Gilmore 1995) and halo (Laird et al. 1988)
populations in the Milky Way. 
}
\end{figure}

\item There is little evidence from the data in Figure 7
for any redshift evolution in the metallicity of DLAs. 
This question has also been addressed using the abundance
of Fe which, thanks to its rich absorption spectrum,
can be followed to higher redshifts and 
lower metallicities than Zn (Prochaska, Gawiser, \& Wolfe 2001). 
Once allowance is 
made for the fraction of Fe in dust grains, 
Vladilo (2002b) finds a gradient 
of $-0.32$ in the linear regression
of [Fe/H] vs. $z_{\rm abs}$.
However, while Figure 7 suggests that
the chances of finding
a DLA with [Zn/H]\,$> -1.0$ are possibly 
greater at $z < 1$, 
the column density-weighted 
metallicities---which measure the density of
metals per comoving volume---are consistent with no evolution 
over the range of redshifts probed so far,
irrespectively of whether Zn or Fe are
considered (see Figure 9). Evidently, the census
of metals at all redshifts is dominated by high column 
density systems of low metallicity.

%
%
\begin{figure} 
\vspace*{-3.25cm} 
\hspace*{-1.0cm}
\epsfig{figure=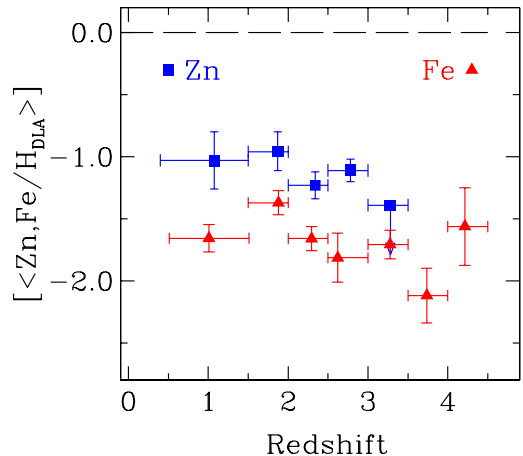,width=13cm,angle=270}
\vspace*{-3.0cm} 
\caption[]{
Column density-weighted metallicities of DLAs in different
redshift intervals, from the surveys by 
Pettini et al. (1999) for Zn, and Prochaska \& Wolfe (2002) for Fe.
The lower abundance of Fe relative to Zn probably reflects
the presence of moderate amounts of dust in most DLAs (Vladilo 2002b).
}
\end{figure}

The lack of evolution in both the neutral gas and metal 
content of DLAs  
was unexpected and calls into question the notion
that these absorbers are unbiased tracers of these 
quantities on a global scale. On the other hand,
the paucity of data at redshifts $z < 1$, that is
over a time interval of 
more than half of the age of the universe (Table 1),
makes it difficult to draw firm conclusions and it may yet
be possible to reconcile existing measurements
with models of cosmic chemical evolution 
(Pei, Fall, \& Hauser 1999; Kulkarni \& Fall 2002).
\end{enumerate}

\subsection{Element Ratios}

So far we have considered only the overall metallicity
of DLAs as measured by the [Zn/H] ratio. However, the 
relative abundances of different elements offer additional
insights into the chemical evolution of this population
of galaxies, as we shall now see. This aspect of the work has really
blossomed with the advent of efficient echelle spectrographs 
on both the Keck and VLT facilities, which have allowed 
the absorption lines of 
a wide variety of elements to be recorded simultaneously,
often with exquisite precision.

\subsubsection{Dust in DLAs}
The presence of dust in DLAs can be inferred by comparing
the gas phase abundances of two elements which in local
interstellar clouds are depleted by differing amounts.
The [Cr/Zn] ratio is one of the most suitable of such
pairs for the reasons described above. It became apparent
from the earliest abundance measurements in DLAs that
this ratio is generally sub-solar, as expected if a fraction
of the Cr has been  incorporated into dust grains.
Figure 10 shows this result for a subset of the DLAs
in Figure 7; similar plots are now available for larger
samples of DLAs and for other pairs of elements, one of 
which is refractory and the other is not (e.g.
Prochaska \& Wolfe 1999; 2002).

From this body of data it is now firmly established that
the depletions of refractory elements are generally lower
in DLAs than in interstellar clouds of similar column density
in the disk of the Milky Way. The reasons for this are not entirely
clear. The question has not yet been addressed quantitatively;
qualitatively the effect is probably related to the lower metallicities
of the DLAs and the likely higher temperature of the interstellar medium
in these absorbers (Wolfire et al. 1995; Petitjean, Srianand, \& Ledoux 2000;
Kanekar \& Chengalur 2001). Figure 10 does seem to indicate a weak
trend of decreasing Cr depletion with decreasing metallicity,
also supported by the results of Prochaska \& Wolfe (2002).

Typically, it is found that refractory elements are depleted
by about a factor of two in DLAs---a straightforward average of
the measurements in Figure 10 yields a
mean $\langle$[Cr/Zn]$\rangle = -0.3^{+0.15}_{-0.2}$~($1 \sigma$ limits).
When we combine this value with the mean metallicity of DLAs,
${\rm [} \langle{\rm Zn/H}_{\rm DLA}\rangle {\rm ]} = -1.13$,
or $\langle Z_{\rm DLA}\rangle  = 1/13\,Z_{\odot}$, 
we reach the conclusion that in damped systems the ``typical''
dust-to-gas ratio is only about $\approx 1/30$ of the Milky Way value
(although there is likely to be a large dispersion
from DLA to DLA, reflecting the range of metallicities
evident in Figure 7).
In the disk of our Galaxy, there a well determined relationship between the 
neutral hydrogen column density and the visual extinction,
$\langle N$(H~I)$\rangle / \langle A_V \rangle = 1.5 \times
10^{21}$~cm$^{-2}$~mag$^{-1}$ (Diplas \& Savage 1994), where $A_V$ is the
extinction (in magnitudes) in the $V$ band. For the typical damped \lya\ system
with neutral hydrogen column density $N$(H~I) = $1 \times 10^{21}$~cm$^{-2}$
and dust-to-gas ratio 1/30 that of the local ISM,
we therefore expect a trifling $A_V \simeq 0.02$~mag in the rest-frame 
$V$ band.  Of more interest is the
far-UV extinction, since this is the spectral region observed at optical
wavelengths at redshifts $z = 2 - 3$. Adopting the SMC
extinction curve (Bouchet et al. 1985)---which may be the appropriate one to
use at the low metallicities of most DLAs---we calculate that a damped
\lya\ system will typically introduce an extinction at 1500 \AA\ 
of $A_{1500} \simeq 0.1$~mag in the spectrum of a background QSO.
Such a small degree of obscuration is consistent
with the mild reddening found in the spectra of QSO with DLAs, 
compared to the average UV continuum slope of QSOs without
(Pei, Fall, \& Bechtold 1991).

%
%
\begin{figure} 
\vspace*{-1.750cm} 
\epsfig{figure=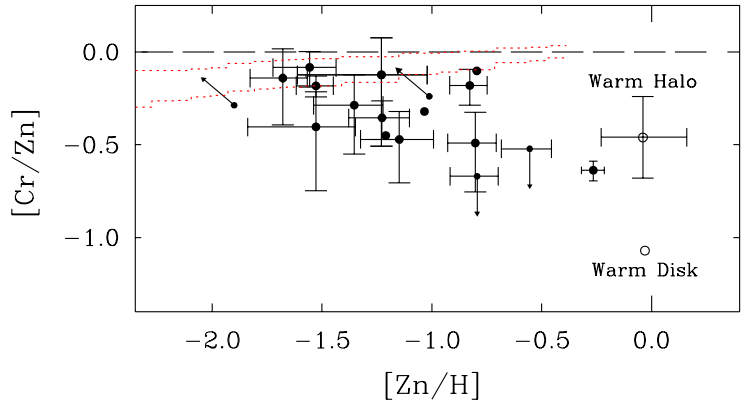,width=12.5cm}
\vspace*{-8.75cm} 
\caption[]{(Reproduced from Pettini et al. 1997a).
Cr abundance relative to Zn in 18 damped \lya\ systems (filled
symbols). The region within the dotted lines 
(reproduced from Ryan, Norris, \& Beers 1996) 
indicates how the [Cr/Fe] ratio varies in 
Galactic stars in this metallicity regime.
The open circles show the typical [Cr/Zn] ratios measured
in interstellar clouds in the disk and halo of our Galaxy,
where the underabundance of Cr relative to Zn is ascribed to
dust depletion (Savage \& Senbach 1996).
} 
\end{figure}

\subsubsection{Alpha-capture elements}

The moderate degree of depletion of refractory elements
in DLAs has motivated a number of attempts to correct
for the fractions missing from the gas phase (Vladilo 2002a
and references therein) and thereby explore {\it intrinsic}
(rather than dust-induced)
departures from solar relative abundances.
The basic idea, which is discussed extensively in the 
contribution to this volume by Francesca Matteucci,
is that different elements are produced by stars of
different masses and therefore different lifetimes.
Thus, the relative abundances of two elements
can, under the right circumstances, provide clues
to the previous star formation history of the galaxy,
or stellar population, under consideration 
(Wheeler, Sneden, \& Truran 1989). Such clues
are not always easy to decipher, however. 
For one thing, they rely on our incomplete, 
and mostly theoretical, knowledge of the 
stellar yields. Secondly, we must assume
a `standard' initial mass function (IMF)
because, if we were free to alter at will
the relative proportions of high and low mass stars,
then we would obviously be able to reproduce most
element ratios, but we could scarcely claim to have learnt
anything in the process. Fortunately, all available
evidence (including that from DLAs, Molaro et al. 2001)
points to a universal IMF as a reasonable first
order approximation (Kennicutt 1998a).

One of the cornerstones of this kind of 
approach is the well established overabundance
of the alpha-capture elements relative
to Iron in metal-poor stars of the Galactic halo.
Mg, Si, Ca, and Ti are generally overabundant by
factors of between two and three in stars where Fe is below
one tenth solar, i.e. [$\alpha$/Fe] = +0.3 to +0.5
when [Fe/H]\,$\simlt -1.0$ 
(Ryan et al. 1996).
This result can be understood if approximately 
two thirds of the Fe (and other Fe-peak) elements
are produced by Type Ia supernovae (SN) 
and released into the ISM 
with a time lag of about  1\,Gyr relative 
the $\alpha$-capture elements (and one third of the Fe)
manufactured by the massive stars which 
explode as Type II supernovae.

%
\begin{figure}  
\hspace{-0.5cm}
\centerline{\includegraphics[width=17pc]{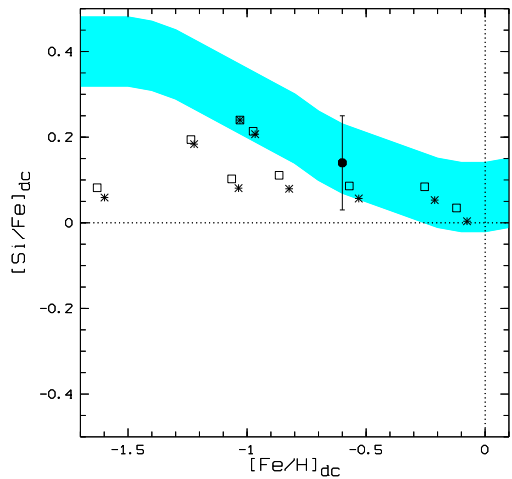}}
\vspace{-3.30cm}
\caption{(Reproduced from Ledoux et al. 2002).
Dust-corrected abundance ratios of Si relative to Fe 
versus DLA metallicity, as measured from the dust-corrected Fe abundances.
Errors are typically $\pm 0.1$\,dex.
Different symbols are used for different dust depletion patterns
adopted when correcting the observed abundances.
The shaded area shows the region occupied by Galactic stars 
in the disk and halo over this range of metallicities.
}
\end{figure}

In this picture, 1\,Gyr is therefore the time over which
the halo of our Galaxy became enriched to a metallicity
[Fe/H]\,$ = -1$, ultimately reflecting the
rate at which star formation proceeded in this stellar component
of the Milky Way. Clearly, the situation could be different
in other environments (Gilmore \& Wyse 1991; Matteucci \& Recchi 2001).
The thick disk, for example, evidently reached solar 
abundances of the $\alpha$-elements in less than 1\,Gyr, since
the $\alpha$ overabundance---or more correctly the Fe 
deficiency---seems to persist to this high 
level of metallicity (Fuhrmann 1998).

Do damped \lya\ systems, which as we have seen
are generally metal-poor, show an overabundance
of the $\alpha$ elements? This question has been 
addressed by several authors and there seems to be
a general consensus that there is not a unique answer.
As can be seen from Figure 11, while some DLAs conform
to the pattern seen in Galactic stars, many others 
do not, in that they exhibit near solar values
of [Si/Fe] even when [Fe/H] is $\ll -1$
(Molaro et al. 2000; Pettini et al. 2000a; Ledoux et al. 2002;
Prochaska \& Wolfe 2002; Vladilo 2002b).
Presumably, these are galaxies where star formation
has proceeded only slowly, or intermittently,
allowing the Fe abundance to `catch up' with that
of the Type II supernova products. The
Magellanic Clouds may be local counterparts of these DLAs
(Pagel \& Tautvaisiene 1998).
Thus, the chemical
clues provided by the these element ratios are another
demonstration, together with the wide range in 
metallicity at the same epoch (\S2.3) and 
the morphologies of the absorbers (\S2.1),
that DLAs trace a diverse population of galaxies,
with different evolutionary histories. Their common
trait is simply a large cross-section on the sky 
at a high surface density of neutral hydrogen.

\subsubsection{The Nucleosynthesis of Nitrogen}

A case of special interest is Nitrogen, whose 
nucleosynthetic origin is a subject
of considerable interest and discussion. There is general
agreement that the main pathway is a six step process in the CN
branch of the CNO cycle which takes place in the stellar H
burning layer, with the net result that $^{14}$N is synthesised
from $^{12}$C and $^{16}$O. The continuing debate, however,
centres on which range of stellar masses is responsible for the
bulk of the nitrogen production.
A comprehensive reappraisal of the problem
was presented by Henry, Edmunds, \& K\"{o}ppen
(2000) who compiled an extensive set of abundance
measurements and computed chemical evolution models using
published yields.  Briefly, nitrogen has both a primary and a
secondary component, depending on whether the seed carbon and
oxygen are those manufactured by the star during helium burning,
or were already present when the star first condensed out of the 
interstellar medium. 

Observational evidence for this dual nature of nitrogen
is provided mainly from measurements of the N and O abundances
in H~II regions. 
(For consistency with other published work, we depart here from the 
notation used throughout the rest of this article,
and use parentheses
to indicate logarithmic ratios of number densities; 
adopting the recent reappraisal of solar photospheric
abundances by Holweger (2001), we have (N/H)$_{\odot} = -4.07$;
(O/H)$_{\odot} = -3.26$; and (N/O)$_{\odot} = -0.81$).
In H~II regions of nearby galaxies, (N/O)
exhibits a strong dependence on (O/H) when the latter is greater
than $\sim 2/5$ solar; this is generally interpreted 
as the regime where secondary N becomes 
dominant.
At low metallicities on the other hand, 
when (O/H)$\,\simlt -4.0$ (that is, $\simlt 1/5$ solar), 
N is mostly primary and tracks O; this results in a 
plateau at (N/O)~$\simeq -1.5$ (see Figure 12).

%
\begin{figure}  
\vspace{-2.0cm}
\centerline{\includegraphics[width=30pc,angle=270]{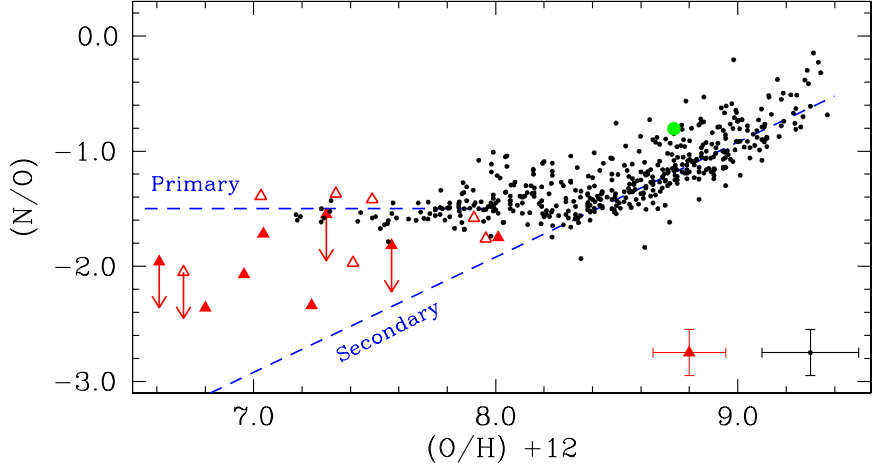}}
\vspace{-2.95cm}
\caption{Abundances of N and O in extragalactic H~II regions
(small dots) and damped \lya\ systems (large triangles).
Sources for the H~II region measurements
are given in Pettini et al. (2002a).
Filled 
triangles denote DLAs where the abundance of O 
could be measured directly, while open triangles
are cases where S was used as a proxy for O.
The error bars in the bottom right-hand corner
give an indication of the typical uncertainties;
the large dot corresponds to the 
solar abundances of N and O from the recent reappraisal
by Holweger (2001).
The dashed lines are approximate
representations of the secondary and primary
levels of N production (see text).
}
\end{figure}

The principal sources of primary N are thought to be intermediate
mass stars, with masses $4 \simlt M/M_{\odot} \simlt 7$, during the 
asymptotic giant branch (AGB) phase. 
A corollary of this hypothesis is that the release
of N into the ISM should lag behind that of O
which, as we have seen, 
is widely believed to be
produced by massive stars which explode 
as Type II supernovae 
soon after an episode of star formation.
Henry et al. (2000) calculated this time delay 
to be approximately 250\,Myr; at low
metallicities the (N/O) ratio could then perhaps be used as a
clock with which to measure the past rate of star formation,
as proposed by Edmunds \& Pagel (1978).
Specifically, in metal-poor galaxies which have 
only recently experienced a 
burst of star formation one may expect to find
values of (N/O) {\it below} the primary plateau
at (N/O)~$\simeq -1.5$, provided the fresh Oxygen 
has been mixed with the ISM 
(Larsen, Sommer-Larsen, \& Pagel 2001).

As pointed out by Pettini, Lipman, \& Hunstead (1995), 
clues to the nucleosynthetic origin 
of nitrogen can also be provided by measurements
of N and O in high redshift DLAs.
Apart from the obvious interest in taking such abundance measurements
to the distant past, when galaxies
were young, one of the advantages of DLAs is that,
thanks to their generally low metallicities, they 
probe a regime where local H~II region abundance measurements are
sparse or non-existent and where the effect of a delayed
production of primary nitrogen should be most pronounced.

Figure 12 shows the most recent compilation of data relevant
to this question. The fact that all DLA measurements fall 
within the region in the
(N/O) vs. (O/H) plot bounded by the primary
and secondary levels of N production
provides empirical evidence in support of 
currently favoured ideas for the nucleosynthesis of 
primary N by intermediate mass stars.
The uniform value (N/O)\,$\simeq -1.5$ 
seen in nearby metal-poor star-forming galaxies
can be understood in this scenario if these
galaxies are not young,
but contain older stellar populations,
as indicated by a number of imaging studies
with {\it HST}.

It is also somewhat 
surprisingly to find such a high proportion (40\%) 
of DLAs which have apparently not
yet attained the full primary level of N enrichment
at (N/O)\,$\simeq -1.5$\,. 
Possibly,  the low metallicity regime---where
the difference between secondary and primary nitrogen enrichment
is most pronounced---preferentially selects
young galaxies which have only recently
condensed out of the intergalactic medium and begun forming
stars. A more speculative alternative, which needs to
be explored computationally, is that at low
metallicities stars with masses lower than 
$4 M_{\odot}$ may make a significant contribution
to the overall N yield (Lattanzio et al., in preparation;
Meynet \& Maeder 2002).
The release of primary N may, under these circumstances,
continue for longer than 250\,Myr, perhaps for a substantial
fraction of the Hubble time at the median $\langle z \rangle = 2.5$
of our sample.

In concluding this section, it is evident that DLAs are a rich
source of information on nucleosynthesis in the early
stages of galaxy formation. Element abundances 
in DLAs are increasingly being taken into consideration,
together with stellar and H~II region data from local
systems, in models of the chemical evolution of 
galaxies and in the calculation of stellar yields.
The chemical clues they provide will be even more valuable
once their connection to today's galaxies in the Hubble
sequence is clarified.

\section{The Lyman Alpha Forest}

The next component to be considered is 
the all-pervading intergalactic medium which manifests 
itself as a multitude of individual \lya\ absorption lines
bluewards of the \lya\ emission line
of every QSO. As can be appreciated from Figures 13 and 14,
the effect is dramatic at high redshift.
Observationally, the term \lya\ forest is used to indicate
absorption lines with column densities in the range
$10^{16}$ \,$\simgt$\,$N{\rm (H I)}$\,$\simgt$\,$10^{12}$ cm$^{-2}$
(see Figure 5).

%
%
\begin{figure}
\centerline{\includegraphics[width=22pc,angle=270]{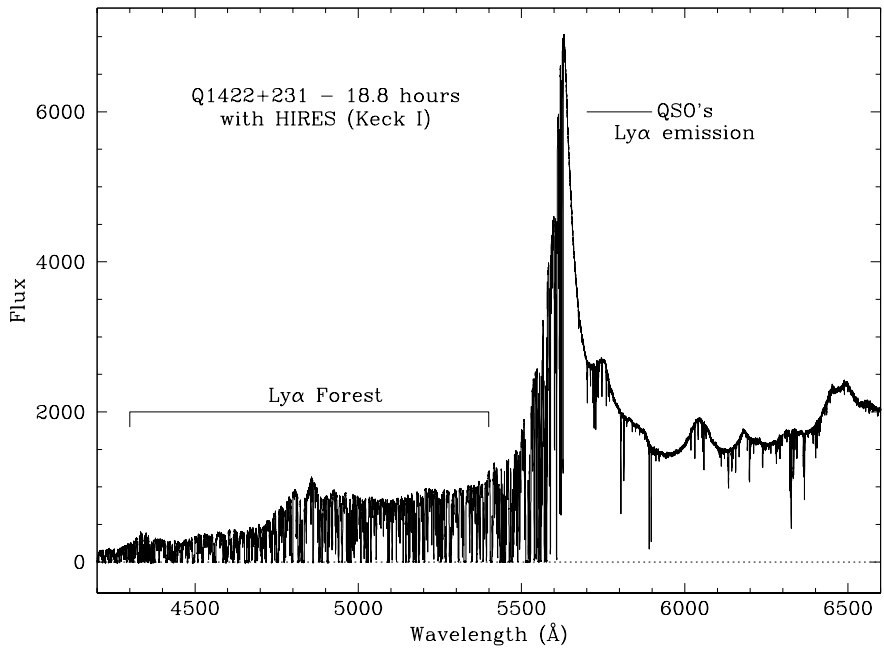}}
\caption{(Reproduced from Ellison 2000). 
This is one of the best QSO spectra ever obtained
thanks to the combination of the bright magnitude of the 
gravitationally lensed QSO Q1422+231 ($V = 16.5$), long exposure
time (amounting to several nights of observation), and
high spectral resolution offered by the Keck echelle
spectrograph (FWHM\,$\simeq 8$\,km~s$^{-1}$). 
The signal-to-noise ratio in the continuum
longwards of the \lya\ emission line is
between 200 and 300. 
At these high redshifts ($z_{\rm em} = 3.625$)
the \lya\ forest eats very significantly into
the QSO spectrum below the \lya\ emission line
and, with the present resolution, 
breaks into hundreds of absorption components
(see Figure 14).
}
\end{figure}

%
%
\begin{figure}
\centerline{\includegraphics[width=30pc]{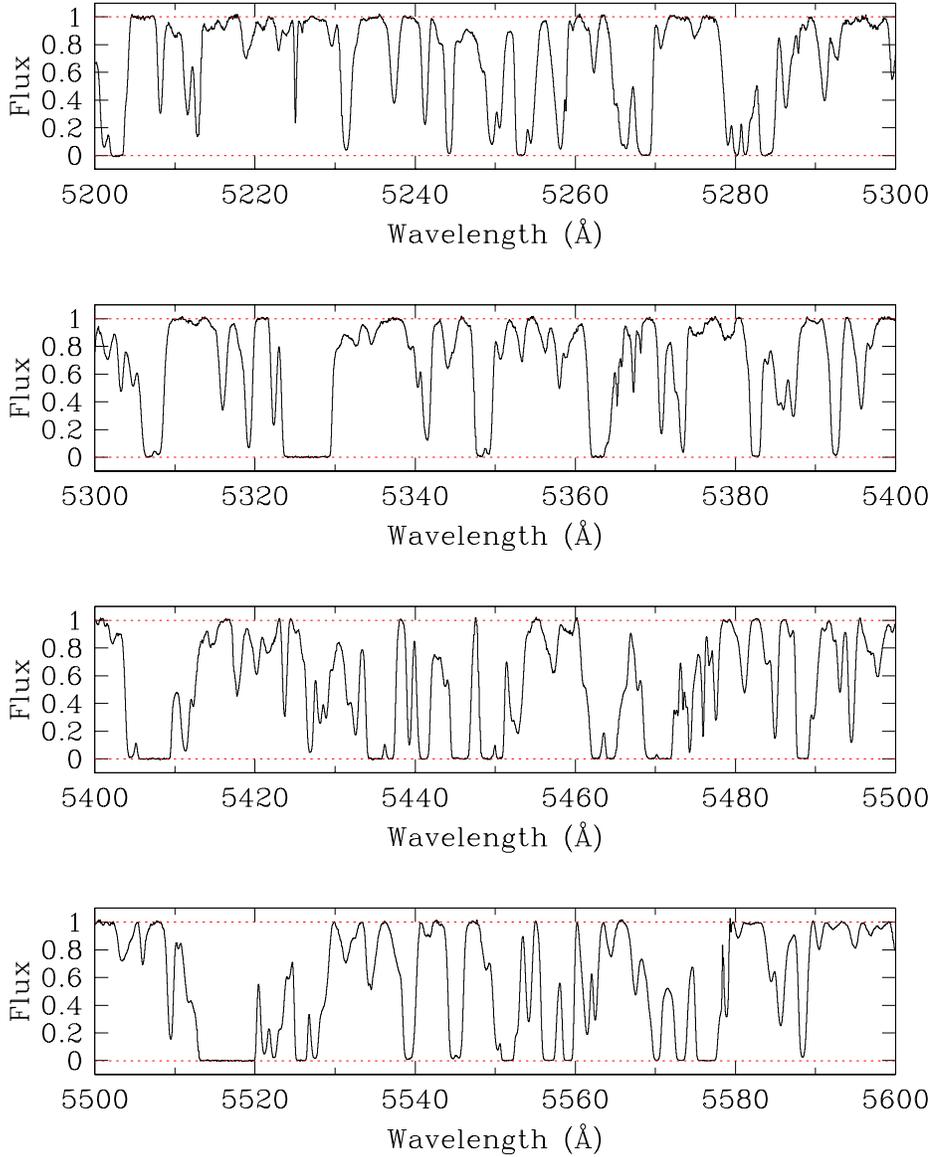}}
\caption{(Reproduced from Ellison 2000). 
Portion of the \lya\ forest between $z_{\rm abs} = 3.277$ 
and 3.607 in the Keck 
spectrum of Q1422+231 shown in Figure 13. There are more than 50
individual absorption components in each 100\,\AA-wide stretch
of spectrum.}
\end{figure}

Hydrodynamical simulations have shown that the \lya\ forest 
is a natural consequence of the formation of large scale structure
in a universe dominated by cold dark matter 
and bathed in a diffuse ionising background 
(see Weinberg, Katz, \& Hernquist 1998 
for an excellent review of the
ideas which have led to this interpretation).
An example is reproduced in Figure 15.
Artificial spectra generated by throwing random sightlines
through such model representations of the high redshift universe
are a remarkably good match to real spectra of the \lya\ forest.
In particular, the simulations are very successful at
reproducing the column density distribution of H~I in Figure 5,
the line widths and profiles, and the evolution 
of the line density with redshift. 
Consequently, much of what we have
learnt about the IGM in the last few years has 
been the result of a very productive interplay between 
observations of increasing precision and simulations
of increasing sophistication. This modern view
of the \lya\ forest is often referred to as 
the `fluctuating Gunn-Peterson' effect.

There are two important properties of the \lya\ forest which we
should keep in mind. One is that it is highly ionised, so that
the H~I we see directly is only a small fraction
($\sim10^{-3}$ to $\sim10^{-6}$) of the total amount of hydrogen
present. With this large ionisation
correction it appears that the forest can account for most of the
baryons at 
high, as well as low, redshift (Rauch 1998; 
Penton, Shull, \& Stocke 2000);
that is $\Omega_{{\rm Ly}\alpha} \approx 0.02~h^{-2}$.
Second, the physics of the absorbing gas is relatively simple 
and the run of optical depth $\tau$(\lya) with redshift 
can be thought of as a `map'
of the density structure of the IGM along a given line of sight.
At low densities, where 
the temperature of the gas is determined by the balance between 
photoionisation heating (produced by the intergalactic ionising
background)
and adiabatic cooling (due to the expansion of the universe), 
$\tau$(\lya)$ \propto {\rm (}1 + \delta{\rm )}^{1.5}$, where  
$\delta$ is the overdensity of baryons
$\delta \equiv {\rm (}\rho_{\rm b}/\langle \rho_{\rm b} \rangle - 1{\rm )}$.
At $z = 3$, $\tau$(\lya)$ = 1$ corresponds to a region of the IGM which 
is just above the average density of the universe at that time
($\delta \approx 0.6$). The last absorption line in the second
panel of Figure 14, near 5395\,\AA, is an example of a
\lya\ line with $\tau \simeq 1$.
The idea that, unlike galaxies,
the forest is an unbiased tracer of mass has prompted,
among other things, attempts to recover the initial spectrum of density fluctuations
from consideration of the spectrum of line optical depths in the
forest (Croft et al. 2002)

%
\begin{figure}  
\vspace{1.0cm}
\centerline{\includegraphics[width=22pc]{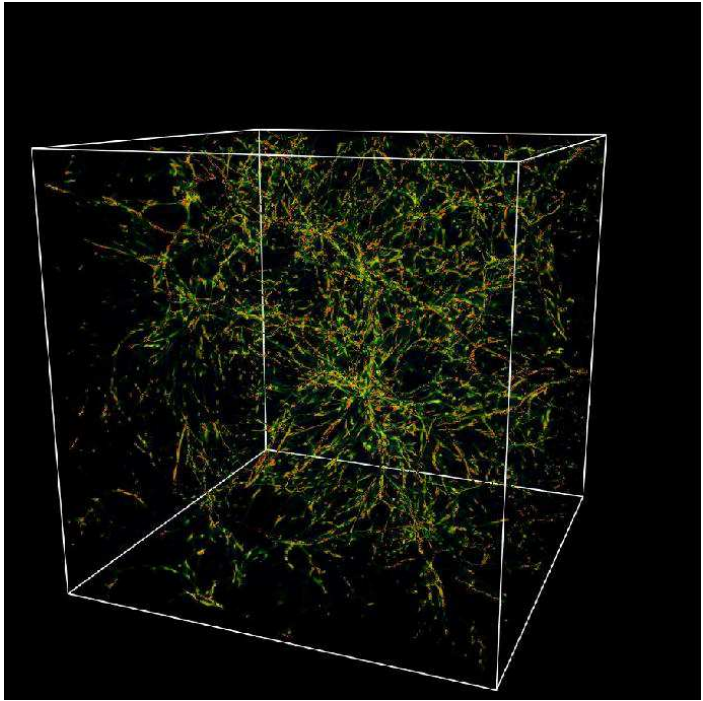}}
\vspace{0.75cm}
\caption{(Reproduced from http://astro.princeton.edu/\~\,cen). Distribution of
neutral gas at $z = 3$ from hydrodynamic cosmological simulation
in a spatially flat, COBE-normalized, cold dark matter model
with the cosmological parameters adopted in this article (\S1.1). The box size
is 25\,Mpc/$h$ (comoving) on the side, and the number of particles
used in the simualation is $768^3$. The structure seen in this
(and other similar simulations) reproduces very well the 
spectral properties of the \lya\ forest 
when artificial spectra are generated along random sightlines through
the box.
}

\end{figure}

\subsection{Metals in the \lya\ Forest}
%
%
\begin{figure}
\centerline{\includegraphics[width=30pc]{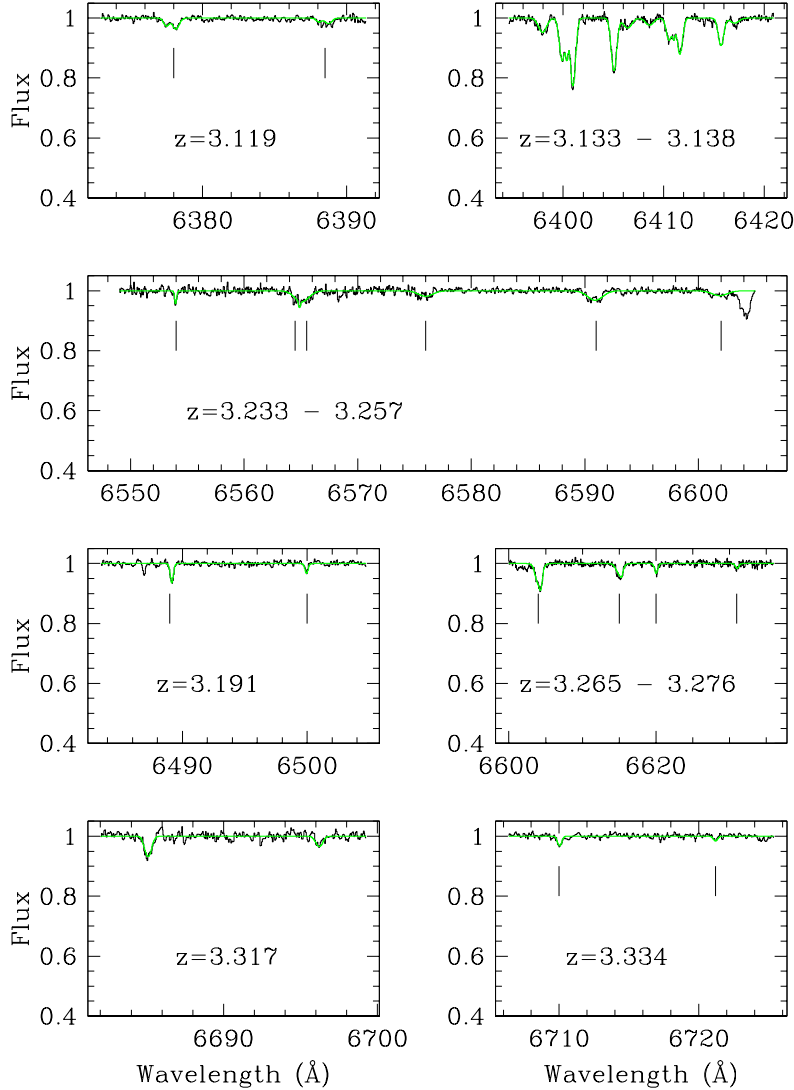}}
\vspace{-1.0cm}
\caption{(Reproduced from Ellison et al. 2000). 
Examples of weak C~IV lines identified in the spectrum
of Q1422+231; most of these would have remained undetected
in spectra of lower signal-to-noise ratios.
Green (grey) lines show the profile fits used to deduce
the column density of C~IV.
The weakest C IV systems are indicated
with tick marks to guide the eye.
}
\end{figure}
The lack of associated metal lines was originally
one of the defining characteristics of the \lya\ 
forest and was interpreted as evidence 
for a primordial origin of the clouds (Sargent et al. 1980).
However, this picture was 
shown to be an oversimplification by the first
observations---using the HIRES spectrograph on the Keck~I
telescope---with sufficient sensitivity to detect the weak
C~IV~$\lambda \lambda 1548, 1550$ doublet associated with \lya\
clouds with column densities log~$N$(H~I)$ \simgt 14.5$
(Cowie et al. 1995; Tytler et al. 1995).
Typical column density ratios in these
clouds are $N$(C~IV)/$N$(H~I)\,$ \simeq 10^{-2}$~--~$10^{-3}$,
indicative of a carbon abundance of about 1/300 of the
solar value, or [C/H]\,$\simeq -2.5$ in the usual notation, 
and with a scatter of perhaps a factor of $\sim 3$
(Dav\'{e} et al. 1998).

The question of interest is `Where do these metals come from?'.
Obviously from stars (we do not know of any other way to produce
carbon!), but are these stars located in the vicinity of the
\lya\ clouds observed---which after all are still at the high
column density end of the distribution of values of $N$(H~I)
for intergalactic absorption---or are we
seeing a more widespread level of metal enrichment, perhaps
associated with the formation of the first stars which
re-ionised the universe at $z > 6$ (Songaila \& Cowie 2002)?

To answer this question we should like to search for metals
in low density regions of the IGM, away from the overdensities
where galaxies form. Observationally, this is a very difficult
task---the associated absorption lines, 
if present at all, would be very weak indeed.
Ellison et al. (1999, 2000) made some progress towards
probing such regions using extremely long exposures with HIRES of two of the
brightest known high-$z$ QSOs, both gravitationally lensed:
APM~08279+5255 and Q1422+231. 
The latter set of data in particular (Figures 13 and 14) 
is of exceptionally high quality, reaching a signal-to-noise ratio S/N
$\simeq 300$ which translates to a limiting rest-frame equivalent
width limit $W_0$($3\sigma$)$\leq 1$~m\AA; this in turn
corresponds to a sensitivity to C~IV absorbers with column
densities as low as $N$(C~IV)~$\simeq 4 \times 10^{11}$~cm$^{-2}$.

And indeed C~IV lines are found at these low levels (see Figure 16),
showing that metals are present in the lowest column
density \lya\ clouds probed, at least down to 
$N$(H~I)$~= 10^{14}$~cm$^{-2}$. 
%
%
\begin{figure}
\centering
\vspace{0.5cm}
\includegraphics[width=.50\textwidth,angle=270]{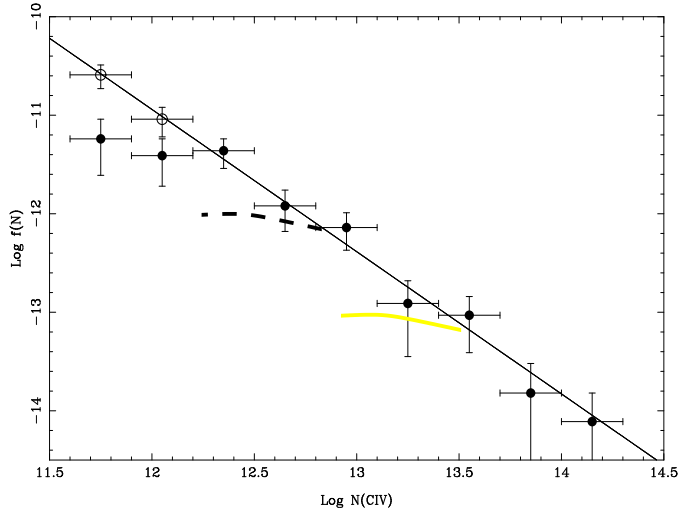}
\vspace{0.3cm}
\caption[]{(Reproduced from Ellison 2000).
C~IV column density distribution in Q1422+231 
at $\langle z \rangle = 3.15$;
$f$($N$) is the number of C~IV systems per column density 
interval and per unit redshift path. 
The filled circles are the data;
the straight line shows the best fitting power-law slope 
$\alpha = 1.44$, assuming the distribution to be of
the form $f{\rm (}N{\rm )} dN = B N^{-\alpha} dN$.
The open circles show the values corrected for incompleteness
at the low column density end; with these correction factors 
there is no indication of a turnover in the column density 
distribution down to the lowest values
of $N$(C IV) reached up to now.
Earlier indications of a turnover shown by the grey
(Petitjean \& Bergeron 1994) and dashed (Songaila 1997)
curves are now seen to be due to the less sensitive
detection limits of those studies, rather than to
a real paucity of weak \lya\ lines.  
}
\end{figure}
As can be seen from Figure 17, the number of weak C~IV lines continues to 
rise as the signal-to-noise ratio of the spectra increases and any 
levelling off in the column density distribution presumably
occurs at $N$(C IV)$ < 5 \times 10^{11}$~cm$^{-2}$. This limit is one 
order of magnitude more sensitive than those reached 
previously.
In other words, we have yet to find any evidence in the \lya\ forest for 
regions of the IGM which are truly of 
primordial composition or have abundances as low as those of
the most metal-poor 
stars in the Milky Way halo.
These conclusions are further supported by the recent detection
of O~VI~$\lambda\lambda 1032,1038$ absorption in the \lya\ forest
at $z = 2$ by Carswell, Schaye, \& Kim (2002).
In agreement with the results of Ellison et al. (2000), these
authors found that most \lya\ forest clouds with 
$N$(H~I)$~ \geq 10^{14}$~cm$^{-2}$ have associated
O~VI absorption and that [O/H] is in the range $-3$ to $-2$.
Weak O~VI lines from regions of lower \lya\ optical depth
have not yet been detected directly, but their presence
is inferred from statistical considerations (Schaye et al. 2000).

\subsection{C~IV at the Highest Redshifts}
%
%
\begin{figure}
\centering
\vspace{-1.0cm}
\hspace{0.5cm}
\includegraphics[width=.650\textwidth,angle=90]{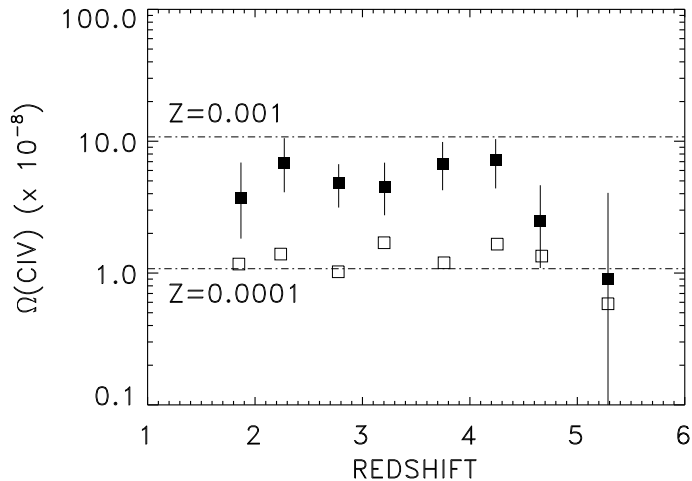}
\vspace{-0.5cm}
\caption[]{(Reproduced from Songaila 2001).
Mass density in C~IV (expressed as fraction of the 
closure density---see eqs. 2.2  and 2.3)
as a function of redshift. The filled symbols
are for C~IV absorption systems in the range
$12 \leq  \log N{\rm (C~IV)} \leq 15$~(cm$^{-2}$).
The dot-dash lines show values of $\Omega ({\rm CIV})$\ computed
assuming $\Omega_b\, h^2 = 0.022, h=0.65$, a C~IV ionisation fraction of 0.5,
and metallicities $Z = 0.0001$\ and $0.001\, Z_{\odot}$, respectively.  
}
\end{figure}
A level of metal enrichment of $10^{-3}$ to $10^{-2}$ of solar in 
regions of the IGM with $N$(H~I)$~ \geq 10^{14}$~cm$^{-2}$
may still be understood in terms of supernova driven winds
from galaxies. The work of Aguirre et al. (2001)
shows that such outflows which, as we shall shortly
see (\S4.5) are observed directly in Lyman break galaxies at $z = 3$,
may propagate out to radii of several hundred kpc before they stall.
However, if O~VI is also present in \lya\ forest clouds
of lower column density, as claimed by Schaye et al. (2000),
an origin in pregalactic stars at much earlier epochs 
is probably required (Madau, Ferrara, \& Rees 2001).

In order to investigate this possibility, Songaila (2001)
extended the search for intergalactic C~IV to $z = 5.5$,
taking advantage of the large number of QSOs with
$z_{\rm em} > 5$ discovered by the Sloan Digital Sky Survey.
The surprising result, reproduced in Figure 18, is that
there seems to be no discernable evolution in the 
integral of the column density distribution of C~IV
from $z = 1.5$\ to $z = 5.5$.  (The reality of a possible
drop in $\Omega ({\rm CIV})$\ beyond $z = 4.5$ 
is questioned by Songaila because incompleteness effects
have not been properly quantified in this difficult
region of the optical spectrum, at $\lambda_{\rm obs} > 8500$\,\AA).
This finding was unexpected and has not yet been properly
assessed. The observed column density of C~IV depends not
only on the overall abundance of Carbon, but also on the shape and
normalisation of the ionising background and on the densities associated with
a given $N$(H~I). Thus, we would have predicted large changes in
$\Omega ({\rm CIV})$ between $z = 5.5$ and 1.5 
in response to the evolving density of ionising sources (QSOs)
and the development of structure in the universe,
even if the metallicity of the IGM had remained constant 
between these two epochs.

Whatever lies behind the apparent lack of 
redshift evolution of $\Omega ({\rm CIV})$,
it is clear that the IGM was enriched with 
the products of stellar nucleosynthesis from the
earliest times we have been able to probe
with QSO absorption line spectroscopy, only
$\sim 1$\,Gyr after the Big Bang.
The measurements of $\Omega ({\rm CIV})$ in Figure 18
suggest a metallicity $Z_{\rm Ly\alpha} \simgt 10^{-3} Z_{\odot}$;
this is a lower limit because 
it assumes that the ionisation of the gas is such that
the ratio C~IV/C$_{\rm tot}$ is near its maximum 
value of about 0.5.
This minimum metallicity can in turn
can be used to infer a minimum number of hydrogen ionising photons
(with energy $h\nu \geq 13.6$\,eV,
corresponding to $\lambda \leq 912$\,\AA) in the IGM, because
the progenitors of the supernovae which 
produce Oxygen, for example,
are the same massive stars 
that emit most of the (stellar) ionising photons.
Assuming a solar relative abundance scale
(i.e. [C/O]\,=\,0), Madau \& Shull (1996)
calculated that the energy of Lyman
continuum photons emitted is 0.2\% of the rest-mass 
energy of the heavy elements produced.\footnote{This is
a lower limit if [C/O]$\, < 0$, as is the case
for low metallicity gas in nearby galaxies 
(e.g. Garnett et al. 1999).}
From this it follows that 
\begin{equation}
\frac{N_{\rm photons}}{N_{\rm baryons}} \times 13.6\,{\rm eV} \simeq 0.002 m_{\rm p} \,c^2 \,Z \simeq 2 \times 10^6\,{\rm eV} \times Z
\end{equation}
(Miralda-Escud\'{e} \& Rees 1997),
where $Z$ is the metallicity (by mass) and $m_{\rm p}$ the mass of the proton.
Since $Z_{\odot} = 0.02$ (Grevesse \& Sauval 1998),
if the \lya\ forest at $z \simeq 5$ had already been enriched to
a metallicity $Z_{\rm Ly\alpha} \simeq 10^{-3}\,Z_{\odot}$,
eq. (3.6) implies that by that epoch stars had emitted approximately
three Lyman continuum (LyC) photons per baryon in the universe.
Whether this photon production is sufficient to have reionised
the IGM by these redshifts depends critically on the unknown
escape fraction of LyC photons from the sites of star formation.

\section{Lyman Break Galaxies}

%
%
\begin{figure}
\vspace{0.5cm}
\hspace{-0.35cm}
\centerline{\includegraphics[width=27.5pc]{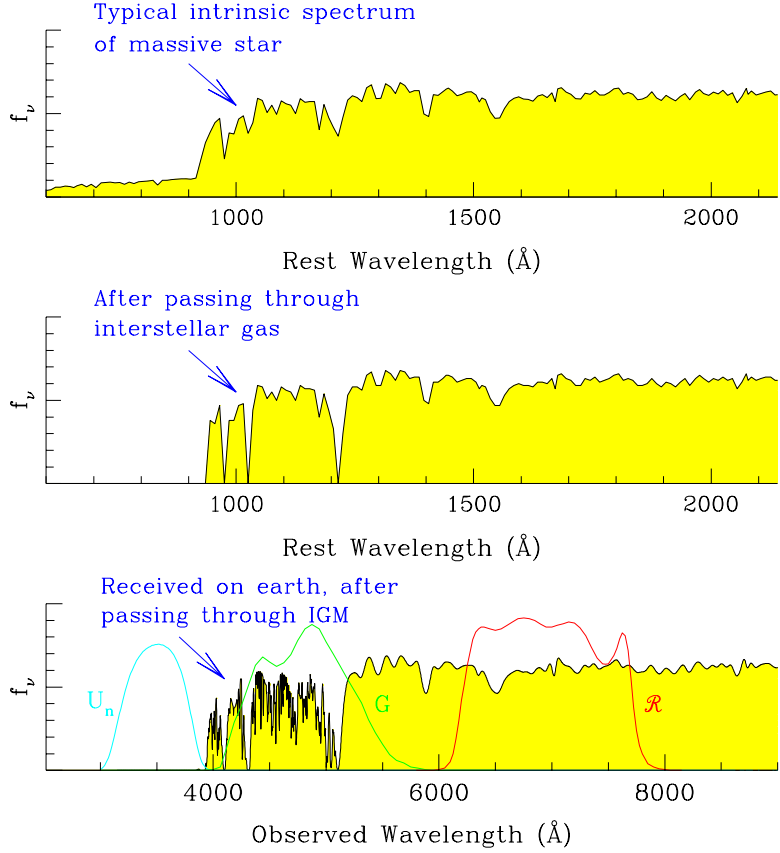}}
\vspace{0.250cm}
\caption{(Courtesy of Kurt Adelberger). 
An illustration of the principles behind the
Lyman break technique. Hot stars have flat
far-UV continua, but emit fewer photons
below 912\,\AA, the limit of the Lyman series
of hydrogen (top panel). These photons are also efficiently
absorbed by any H~I associated with the sites
of star formation (middle panel) and have a short mean free
path---typically only $\sim 40$\,\AA---in the IGM at $z = 3$.
Consequently, when observed from Earth (bottom panel),
the spectrum of a star forming galaxy at $z \simeq 3$
exhibits a marked `break' near 4000\,\AA. With appropriately chosen
broad-band filters, this spectral discontinuity
gives rise to characteristic colours; objects
at these redshifts appear blue in ($G - {\cal R}$)
and red in ($U_n - G$). For this reason, such 
galaxies are sometimes referred to as $U$-dropouts.
A more quantitative description of the Lyman break technique
can be found in Steidel, Pettini, \& Hamilton (1995).
}
\end{figure}

%
%
\begin{figure} 
\vspace*{0.5cm} 
\hspace*{2.5cm}
\psfig{figure=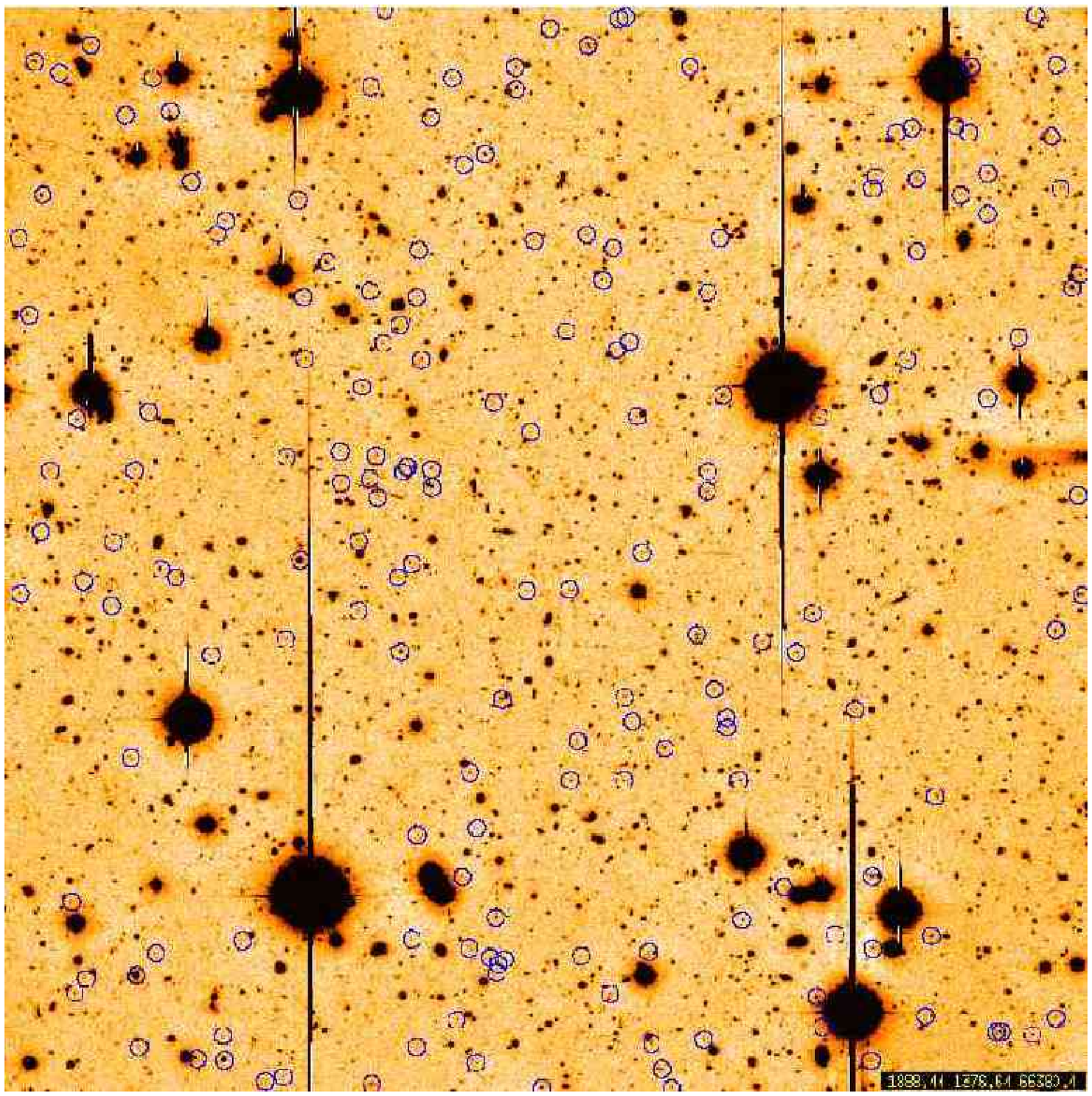,width=8cm} 
\vspace*{0.4cm} 
\caption[]{A typical deep CCD image recorded at the 
prime focus of a 4m-class telescope. This particular
image (of the field designated DSF2237b)
was obtained with the COSMIC camera
of the Palomar Hale telescope, by exposing for a total
of two hours through a custom made ${\cal R}$ filter.
In the $9 \times 9$\,arcmin field of view
(corresponding to co-moving linear
dimensions of $11.6 \times 11.6\,h^{-1}$\,Mpc
at $z = 3$) there are $\sim 3300$
galaxies brighter than ${\cal R} = 25.5$;
140 of these (circled) show Lyman breaks
which place them  
at redshifts between $z = 2.6$ and 3.4\,.
}
\end{figure}

Returning to Figure 1, it must be remembered that 
everything we have learnt on the distant
universe up to this point required
`decoding' the information `encrypted'
in the absorption spectra of QSOs.
It is easy to appreciate, therefore, the 
strong incentive which motivated astronomers
in the 1990s to detect high resdshift galaxies directly.
After many years of fruitless searches, 
we have witnessed since 1995 a 
veritable explosion of data from the 
{\it Hubble Deep Fields} and ground-based surveys
with large telescopes. The turning point
was the realisation of the effectiveness
of the Lyman break technique (Figure 19) in preselecting
candidate $z \simeq 3$ galaxies (Steidel et al. 1996).
Although these galaxies 
constitute only $\sim 3-4$\% 
of the thousands of faint objects, at all redshifts,
revealed by a moderately deep CCD exposure 
at the prime focus of a 4\,m telescope (see Figure 20),
they can be readily distinguished on the basis
of their colours alone, when observed through appropriately
selected filters (Figure 21).

%
%
\begin{figure}
\centering
\vspace*{5pt}
\hspace*{-0.25cm}
\psfig{figure=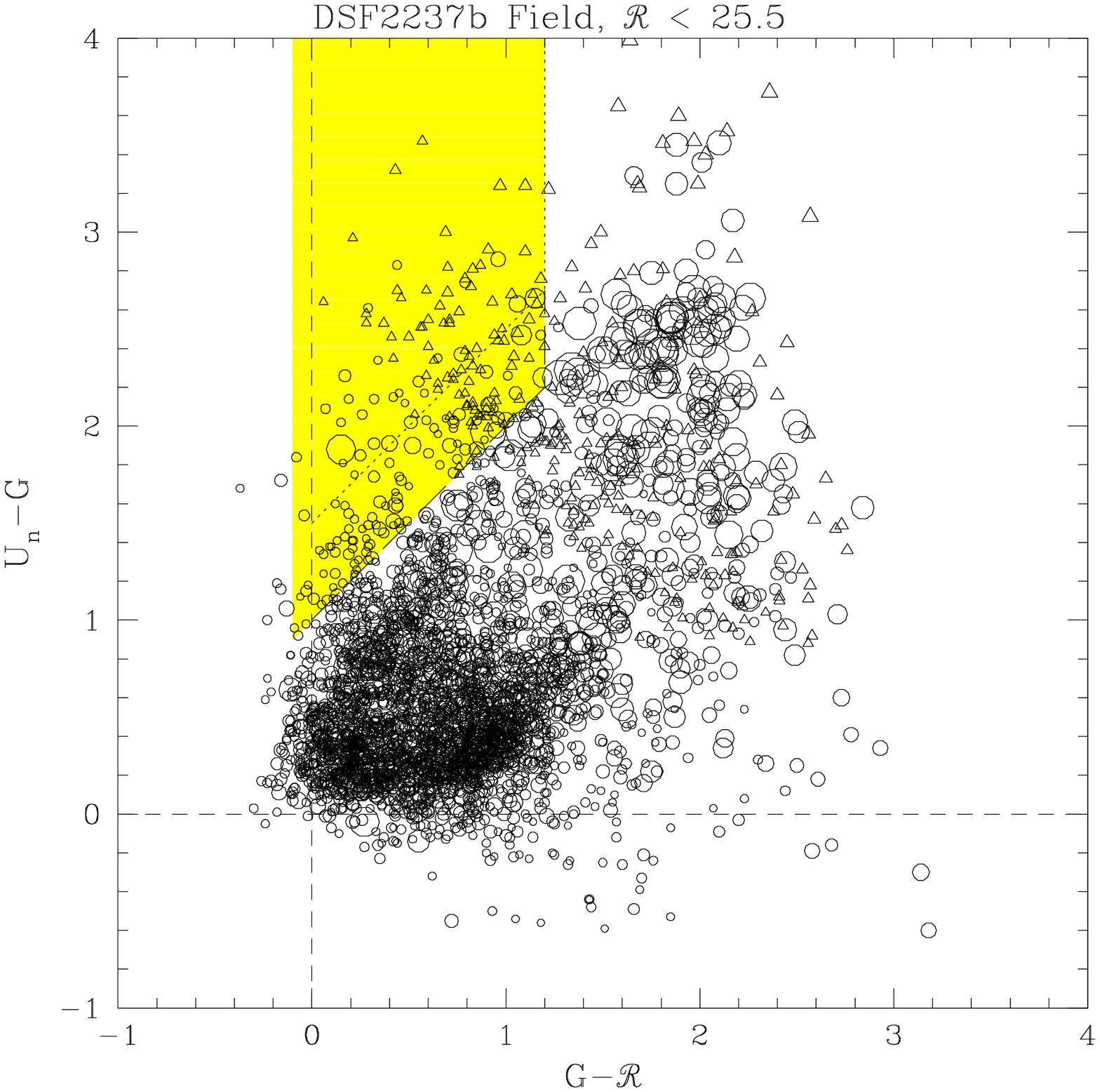,width=7cm} 
\caption{
All the $\sim 3300$ galaxies from Figure 20 are included in this
colour--colour plot. The shaded region shows how the 140
candidate Lyman break galaxies are selected for subsequent spectroscopic
follow-up. The symbol size is proportional to the object magnitude;
circles denote objects detected in all three bands, while
triangles are lower limits in ($U_n - G$) for $U_n$ dropouts.
}
\end{figure}
%
%
\begin{figure}
\centering
\vspace*{0.5cm}
\hspace*{-0.25cm}
\psfig{figure=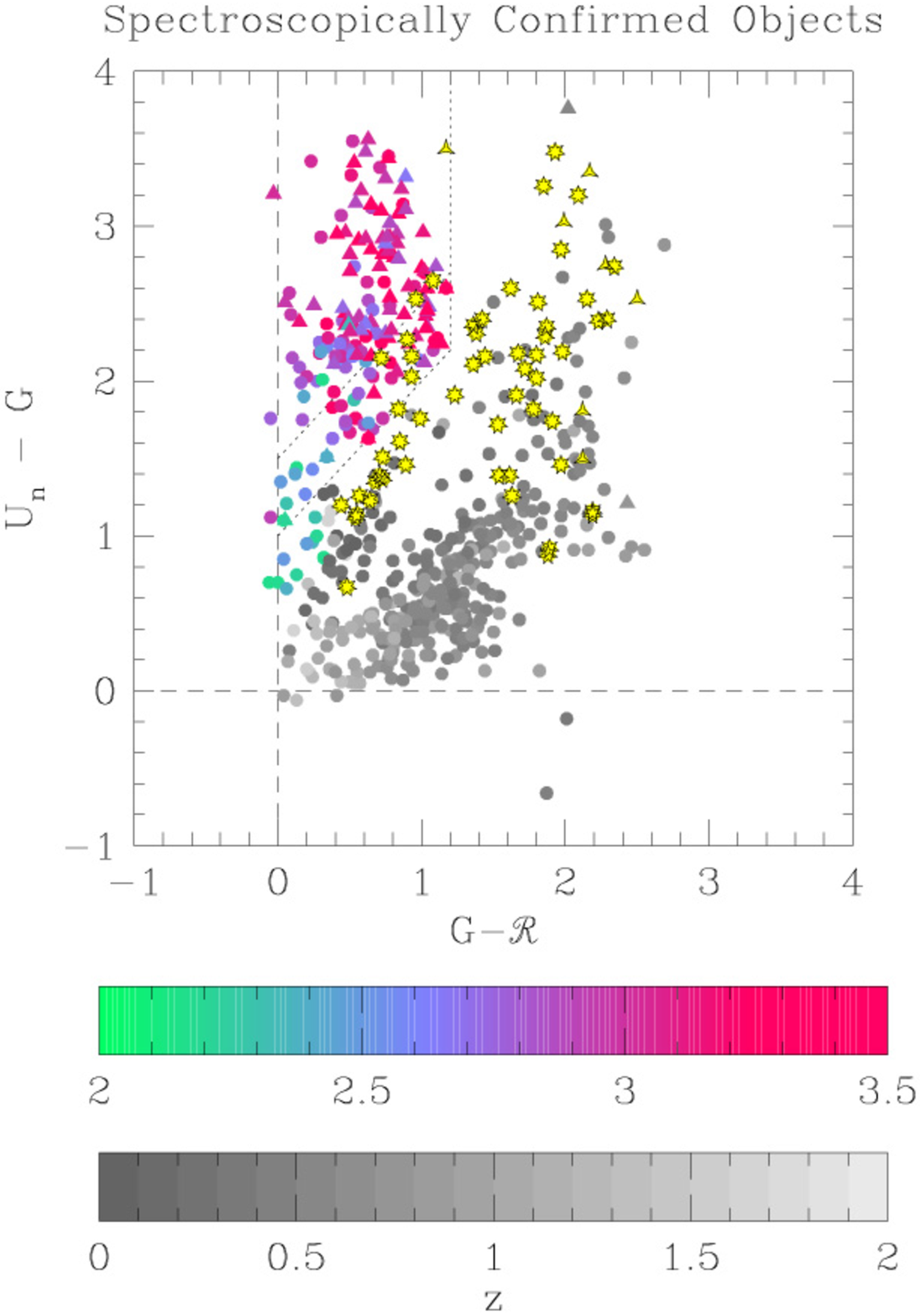,width=10cm} 
\vspace{1.5cm}
\caption{
The location of spectroscopically confirmed
galaxies  (from the surveys by Steidel and collaborators)
on the ($U_n - G$) vs. ($G - {\cal R}$) plot. Triangles
denote objects undetected in the $U_n$ band; stellar
symbols are used for Galactic stars.
}
\end{figure}

The Lyman break technique has been very successful
at finding high redshift galaxies thanks to the combination
of the increasingly large and UV sensitive CCDs used
to identify candidates on the one hand,
and the multi-object spectroscopic capabilities of 
large telescopes required for follow-up and 
confirmation on the other. Thus, samples
of spectroscopically confirmed $z \simeq 3$ galaxies
have grown from zero to more than one thousand in the
space of only five years. As can be seen from Figure 22,
the spectroscopic redshifts generally conform to 
expectations based on just two colours.
Such large samples 
have made it possible to trace the star formation history of the 
universe over most of the Hubble time and to measure the large-scale properties
of this population of galaxies, most notably their clustering and luminosity 
functions (see Steidel 2000 for a review).

In parallel with this work on the Lyman break population
as a whole, in the last few years 
we have also begun to study in more detail
the physical properties of some of the brighter
galaxies in the sample.
The questions which we would like to address are:
\begin{enumerate}
\item What are the stellar populations of the Lyman break galaxies?
\item What are their ages and masses?
\item What are their levels of metal enrichment? 
\item What are the effects of star formation at high $z$
on the galaxies and the surrounding IGM?
\end{enumerate}
Many of these questions link
observational cosmology with stellar and interstellar
astrophysics, and this will become evident as we now explore them 
in turn.

\subsection{Stellar Populations and the Initial Mass Function}
Among the thousand or so known Lyman break galaxies (LBGs),
one, designated MS 1512$-$cB58 (or cB58 for short) 
has provided data of exceptional quality, thanks to
its gravitationally lensed nature.
Discovered by Yee et al. (1996), 
cB58 is, as far as we can tell, a
typical $\sim L^{\ast}$ galaxy at a redshift $z = 2.7276$
magnified by a factor of $\sim 30$
by the foreground cluster MS~1512$+$36 at $z = 0.37$
(Seitz et al. 1998). This fortuitous alignment
makes it by far the brightest known member of the
LBG population and has motivated
a number of studies 
from mm to X-ray wavelengths.

%
%
\begin{figure}
\vspace{-2.50cm}
\hspace{0.2500cm}
\psfig{figure=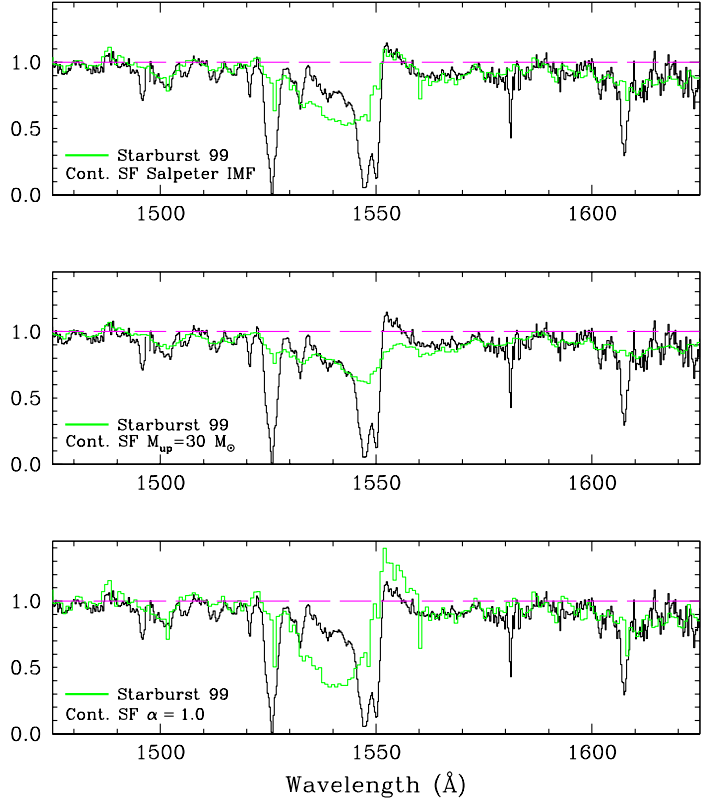,width=12.5cm}
\vspace*{-2.75cm}
\caption[]{Comparisons between {\it Starburst99} (Leitherer et al. 1999) 
population synthesis models with different IMFs (green lines)
and the Keck spectrum of MS~1512-cB58 analysed by
Pettini et al. (2000b) in the region
near the C~IV doublet (black histogram).
The $y$-axis is residual intensity.}
\end{figure}

When we record the spectrum of a $z = 3$ galaxy at 
optical wavelengths, we are in fact observing the 
redshifted far-UV light produced by a whole population
of O and early B stars.  
Such spectra are most effectively analysed with 
population synthesis models, 
the most sophisticated of which is {\it Starburst99} developed by the 
Baltimore group (Leitherer et al. 1999). In Figure 23 we compare 
{\it Starburst99}\/ model predictions 
for different IMFs with a portion of the moderate resolution Keck LRIS spectrum of 
cB58 obtained by Pettini et al. (2000b),
encompassing the C~IV~$\lambda\lambda 1548, 1550$ doublet.

It is important to realise
that the comparison only refers to {\it stellar} spectral features
and does not include the {\it interstellar} lines, 
readily recognisable by their narrower widths 
(the interstellar lines are much stronger in cB58, where we 
sample the whole ISM of the galaxy, than in the models which are based on
libraries of nearby Galactic O and B stars). With this clarification,
it is evident from Figure 23 that the spectral properties of at least 
this Lyman break galaxy are remarkably similar to those of present-day 
starbursts---a continuous star formation model with a Salpeter IMF
provides a very good fit to the observations.
In particular, the P-Cygni profiles of C~IV, Si~IV and N~V are
sensitive to the slope and upper mass limit of the IMF; the best fit in 
cB58 is obtained with a standard Salpeter IMF with slope $\alpha = 2.35$
and $M_{\rm up} = 100 M_{\odot}$ (top panel of Figure 23). 
IMFs either lacking in the most massive stars
or, conversely, top-heavy seem to be excluded by the data
(middle and bottom panels of Figure 23 respectively).

The only significant difference between the observed and synthesised
spectra is in the optical depth of the P-Cygni absorption trough 
which is lower than predicted (top panel of Figure 23). 
A possible explanation is that this is 
an abundance effect---the strengths of the wind lines 
are known to be sensitive to the metallicity of the stars,
since it is through absorption and scattering of photons
in the metal lines that momentum is transferred 
to the gas and a wind is generated.
The {\it Starburst99}\/ spectra shown in Figure 23
are for solar metallicity, but a recent update (Leitherer et al. 2001)
now includes stellar libraries compiled with {\it HST}
observations of hot stars in the Large and Small Magellanic 
Clouds, taken to be representative of a metallicity
$Z \simeq 1/3 Z_{\odot}$.

When these are compared with a higher resolution spectrum
of cB58, obtained with the new Echelle Spectrograph and Imager
(ESI) on the Keck~II telescope, the match to the observed spectrum 
is improved (Figure 24). 
The emission component
of the P-Cygni profile is perhaps underestimated by the
model, but we suspect that there may be additional nebular
C~IV emission from the H~II regions in the galaxy, superposed
on the broader stellar P-Cygni emission. 
We take the good agreement in Figure 24 as an indication that the
young stellar population of cB58 has reached a metallicity
comparable to that of the Magellanic Clouds. This conclusion
is reinforced by measurements of element abundances 
in the interstellar gas, as we shall see in 
\S4.2 below.

Before leaving this topic, it is worth noting 
that with a modest amount of effort (and luck in the 
form of gravitational lensing) it is now possible to obtain
spectra of high redshift galaxies 
of sufficient quality to distinguish 
between the OB stellar populations of the Milky Way and
of the Magellanic Clouds.
As a matter of fact, it is evident from Figure 24
that the resolution of the ESI spectrum of cB58 is superior
to those of Magellanic Cloud stars (and of nearby starburst galaxies)
with which we would like to compare it! 

%
%
\begin{figure}
\centering
\vspace*{-0.75cm}
\hspace*{0.5cm}
\psfig{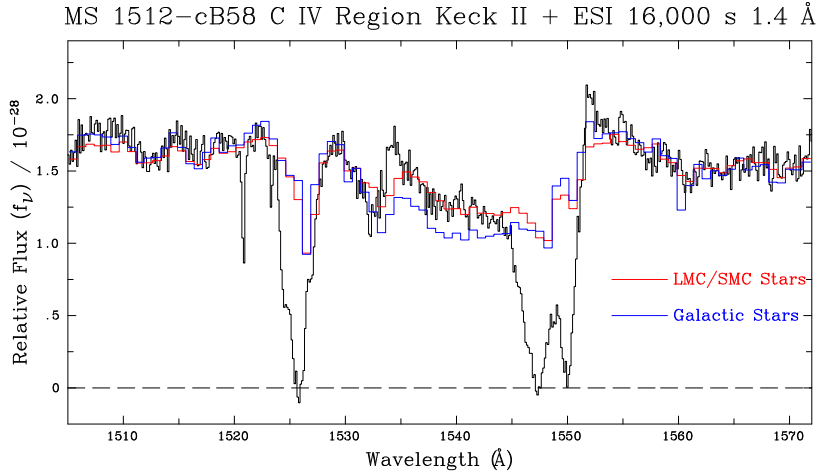}
\vspace{-0.75cm}
\caption[]{A portion of the high resolution spectrum of MS~1512-cB58
obtained by Pettini et al. (2002b), encompassing the C~IV~$\lambda\lambda 1548, 1550$
doublet lines, is compared with {\it Starburst99} synthetic spectra for solar
and $\sim 1/3$ solar (LMC/SMC) metallicities (Leitherer et al. 2001).
}
\end{figure}

\subsection{Element Abundances in the Interstellar Gas}

The ESI spectrum of cB58, which covers the wavelength
region from 1075 to 2800\,\AA\ at a resolution of 58\,km~s$^{-1}$,
is a real treasure trove of information on this galaxy.
For example, it includes 48 interstellar absorption lines 
of elements from H to Zn in a variety of ionisation stages,
from neutral (H~I, C~I, O~I, N~I) to highly ionised
species (Si~IV, C~IV, N~V). The lines are fully resolved
so that column densities can be derived from the analysis of their
profiles. From these data Pettini et al. (2002b) were able
to piece together for the first time a 
comprehensive picture 
of the chemical composition of the interstellar gas
in a Lyman break galaxy
(Figure 25) and examine the clues it provides
on its evolutionary status and 
past history of star formation.

As can be seen from Figure 25, 
the ambient interstellar medium of cB58 is highly
enriched in the elements released by Type~II
supernovae; O, Mg, Si, P, and S all have abundances
of $\sim 2/5$ solar. Thus, even at this relatively
early epoch ($z = 2.7276$ corresponds to 2.5\,Gyr after
the Big Bang in our adopted cosmology---see Table 1), 
this galaxy had already processed more than one third 
of its gas into stars.

%
%
\begin{figure} 
\vspace*{-1.75cm} 
\hspace*{-0.755cm}
\psfig{figure=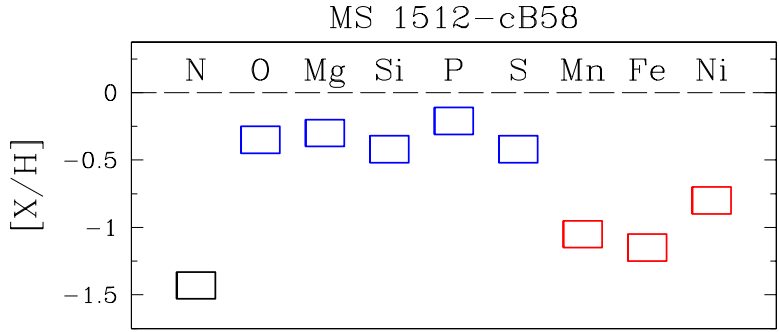,width=10cm,angle=270} 
\vspace*{-2.75cm} 
\caption[]{Pattern of chemical abundances in the ambient
interstellar medium of cB58 deduced by Pettini et al. (2002b).
The vertical height of the 
boxes shows the typical uncertainty in the abundance determinations.
Blue boxes denote elements thought to be 
synthesised by massive stars which explode as 
Type~II supernovae, while red boxes are for the Fe-peak
elements predominantly produced by 
Type~Ia SN. Their release into the ISM, as well as that of N
from intermediate mass stars, lags behind that 
of the Type~II SN products by several 100\,Myr.
} 
\end{figure}

Furthermore, cB58 appears to be chemically young, in that
it is relatively deficient in elements produced by stars
of intermediate and low mass with longer lifetimes
than those of Type~II SN progenitors. 
N and the Fe-peak elements
we observe (Mn, Fe, and Ni) are all less abundant
than expected by factors of between 0.4 and 0.75\,dex.
Depletion onto dust, which is known to be present in cB58,
probably accounts for some of the Fe-peak element
underabundances, but this is not likely to be an important
effect for N. On the basis of current ideas of the nucleosynthesis 
of N, discussed in \S2.4.3, it would appear 
that much of the ISM enrichment in cB58
has taken place within the last 250\,Myr, the lifetime
of the intermediate mass stars believed to be the
main source of N. For comparison, the starburst episode
responsible for the UV and optical light we see
is estimated to be younger than $\sim 35$\,Myr,
on the basis of theoretical models of the
spectral energy distribution at these wavelengths
(Ellingson et al. 1996).

Taken together, these two findings are highly
suggestive of a galaxy caught in the act of converting
its interstellar medium into stars on a few dynamical
timescales---quite possibly in cB58 we are witnessing the 
formation of a galactic bulge or an elliptical galaxy.
The results of the chemical analysis are consistent
with the scenario proposed by Shapley et al. (2001),
whereby galaxies whose UV spectra are dominated by
strong, blueshifted absorption lines, as is the case 
here, are the youngest in the range of ages of LBGs.
These findings also lend support to models
of structure formation which predict that, even at $z \simeq 3$,
near-solar metallicities should in fact be common 
in galaxies with masses greater than $\sim10^{10} M_{\odot}$
(e.g. Nagamine et al. 2001).
The baryonic mass of cB58 is deduced to be
$m_{\rm baryons} \simeq 1 \times 10^{10} M_{\odot}$,
from consideration of its star formation history,
metallicity, and the velocity dispersion of its ionised gas.

\subsection{The Oxygen Abundance in H~II Regions}

How typical are these results of the Lyman break galaxy
population as a whole?
In the nearby universe, element abundances 
in star forming regions have traditionally
been measured from the ratios of optical emission lines
from H~II regions. At $z = 3$ these features move to near-infrared 
(IR) wavelengths and have only become accessible in the last 
two years with the commissioning of high resolution spectrographs
on the VLT (ISAAC) and Keck telescopes (NIRSPEC).
Using these facilities, our group 
has recently completed the first 
spectroscopic survey of Lyman break galaxies in the near-IR, 
bringing together data for 19 LBGs; the galaxies
are drawn from the bright end of the luminosity function, from 
$\sim L^{\ast}$ to $\sim 4\,L^{\ast}$ (Pettini et al. 2001).
Figure 26 shows an example of the quality of spectra which can be secured 
with a 2--3 hour integration.
In five cases 
we attempted to deduce values of the abundance of oxygen
by applying the familiar 
$R_{23}$\,=\,([O~II]\,+\,[O~III])/\Hb\ method first proposed
by Pagel et al. (1979).
We found that generally there remains a significant 
uncertainty, by up to 1\,dex, in the value of (O/H)
because of the double-valued nature of the 
$R_{23}$ calibrator (see Figure 27). 
%
%
\begin{figure}
\vspace*{-2.70cm}
\hspace*{-1.250cm}
\centering
\includegraphics[width=0.85\textwidth,angle=270]{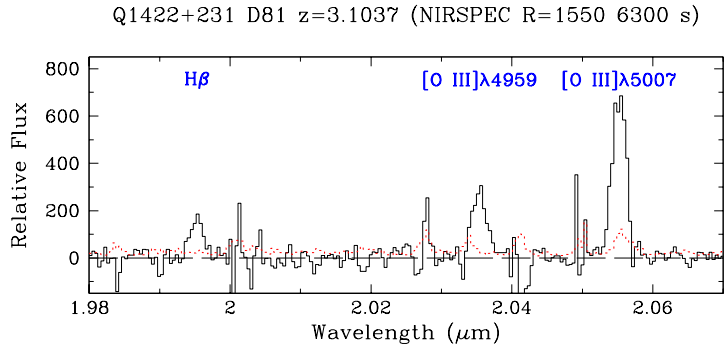}
\vspace*{-3.5cm}
\caption[]{
Example of a NIRSPEC $K$-band spectrum of a Lyman break galaxy
from the survey by Pettini et al. (2001).
The objects targeted typically have $K = 21$ (on the Vega 
scale) and remain undetected in the continuum. However,  the nebular     
emission lines of {\rm [O~III]}\,$\lambda\lambda 4859, 5007$,
{\rm [O~II]}\,$3727$ (not shown), and \Hb\ usually show up clearly 
with exposure times of 2--3 hours. The dotted line is 
the $1 \sigma$ error spectrum.}
\label{}
\end{figure}

%
%
\begin{figure} 
\hspace*{1.25cm}
\epsfig{figure=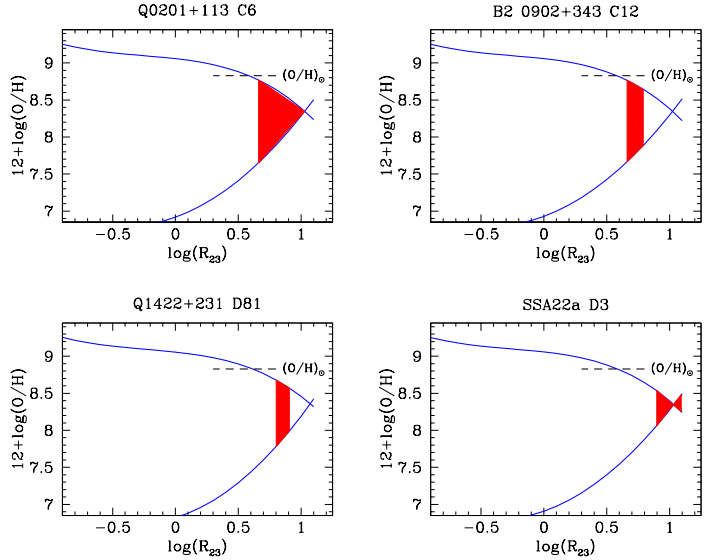,width=8cm,angle=270} 
\caption[]{Oxygen abundance from the 
$R_{23}$\,=\,{\rm ([O~II]+[O~III])}/\Hb\ ratio.
In each panel the continuous lines are the calibration by McGaugh (1991)
for the ionisation index $O_{32}$\,=\,{\rm [O~III]/[O~II]} appropriate to
that object. The shaded area shows the values allowed
by the measured $R_{23}$ and its statistical $1 \sigma$ error.
The broken horizontal line gives for reference the 
solar abundance 12\,+\,log(O/H) = 8.83 from the compilation by Grevesse \& 
Sauval (1998); the recent revision by Holweger (2001) would
bring the line down by 0.09 dex.
}
\end{figure}

Thus, in the galaxies observed,
oxygen could be as abundant as in the interstellar medium 
near the Sun or as low as $\sim 1/10$ solar.
When the $R_{23}$ method is applied to cB58, a similar
ambiguity obtains (Teplitz et al. 2000). The results from
the analysis of the interstellar absorption lines described above (\S4.2)
resolve the issue by showing that the upper branch solution
is favoured (we have no reason to suspect that
the neutral and ionised ISM have widely 
different abundances). It remains to be established whether 
this is also the case for other LBGs.

In the near future this work will shift to 
lower and more easily accessible redshifts
near $z =2.2$ where all the lines of interest,
from [O~II]~$\lambda 3727$ to H$\alpha$, fall 
in near-IR atmospheric transmission windows.
Nevertheless, the determination of element
abundances from nebular emission lines 
will remain a time consuming task
until multi-object spectrographs operating at near-IR 
wavelengths become available on large telescopes,
or until the launch of the 
{\it Next Generation Space Telescope}
(Kennicutt 1998b).

\subsection{Dating the Star Formation Activity}

%
%
\begin{figure} 
\hspace*{2.75cm}
\epsfig{figure=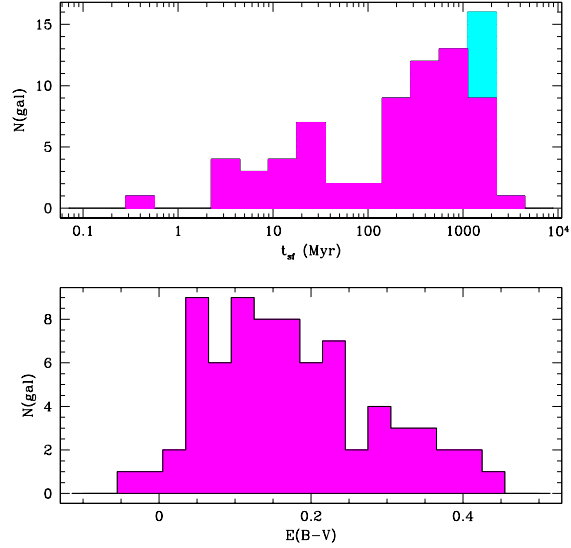,width=7.95cm}
\caption[]{Histograms of best-fitting ages and reddening
for the sample of 81 $z \simeq 3$ LBGs analysed by Shapley et al. (2001).
There is a large spread of ages in the population;
the median age is 320\,Myr and 20\% of the objects
are older than 1\,Gyr. 
The cyan (light grey) bin
corresponds to galaxies with inferred ages 
older than the age of the universe at their redshifts
(an indication of the approximate nature of the
ages derived by SED fitting). 
The median $E$($B-V$)\,$ = 0.155$
for the sample corresponds to attenuations by factors of $\sim 4.5$
and $\sim 2$ at 1600\,\AA\, and 5500\,\AA\ respectively.
}
\end{figure}
%
%
\begin{figure} 
\vspace*{0.55cm} 
\hspace*{2.75cm}
\epsfig{figure=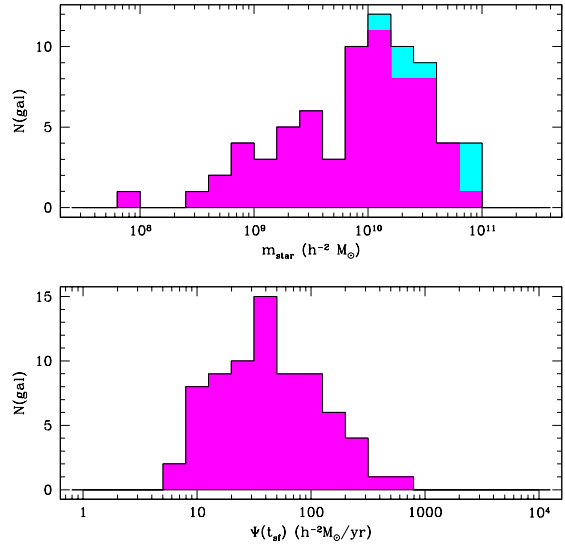,width=7.95cm}
\caption[]{
Histograms of assembled stellar mass and star formation rates
from Shapley et al. (2001). 
By redshift
$z \simeq 3$ a significant fraction of LBGs seem to be 
approaching the stellar mass of today's $L^{\ast}$ galaxies,
$m_{\rm star} \simeq 4 \times 10^{10} M_{\odot}$.
}
\end{figure}

Realistically, the detailed  spectroscopic analysis 
described above can only be applied to a subset 
of LBGs, at the bright end of the luminosity funtion.
However, the coarser 
spectral energy distribution (SED) of Lyman break galaxies
still holds important
information on the star formation episodes.
Broad-band photometry in the optical and near-infrared,
spanning the wavelength interval 900--5500\,\AA\ in the 
rest-frame, is now available for more than one hundred
galaxies at $z \simeq 3$ (Papovich, Dickinson, \& Ferguson 2001;
Shapley et al. 2001). The colours over this range (typically
four colours are used in the analysis) depend on the degree
of dust reddening, $E{\rm (}B-V{\rm )}$,
and on the age of the stellar population, $t_{\rm sf}$.
The two can be decoupled with some degree of confidence 
provided that the SED includes the age-sensitive Balmer break 
near 3650\,\AA, which at $z = 3$ falls between the $H$ and $K$ 
bands---hence the need for accurate near-IR photometry.
A third parameter, the instantaneous star formation rate,
$\Psi$($t_{\rm sf}$), determines the normalisation 
(rather than the shape) of the SED. 
The analyses by Papovich et al. (2001) and 
Shapley et al. (2001) deduced
the best-fitting values of $E{\rm (}B-V{\rm )}$,
$t_{\rm sf}$, and $\Psi$($t_{\rm sf}$) by $\chi^2$
minimisation of the differences between the observed
SEDs and those predicted by the widely used
population synthesis code of Bruzual \& Charlot (1993 and 
subsequent updates).
The results have turned out to be very 
interesting---some would say surprising
(see Figures 28 and 29).

Evidently, Lyman break galaxies span a wide range of ages.
One fifth of the sample considered by Shapley et al. (2001)
consists of objects which apparently have just collapsed
and are forming stars on a dynamical timescale 
($\sim 35$\,Myr). As we have seen, cB58 seems to belong
to this class.
At the other end of the scale, some 20\% of the galaxies
at $z = 3$ have been forming stars for more than 1\,Gyr, placing
the onset of star formation at much higher redshifts ($z > 5-10$).
Furthermore, there appears to be a correlation between
age and star formation rate, with the younger objects
typically forming stars at about ten times the rate
of the older ones and being more reddened on average.
The mean SFR is $\langle \Psi$($t_{\rm sf}$)$\rangle
= 210 h^{-2} \, M_{\odot}$~yr$^{-1}$ for galaxies
with $t_{\rm sf} < 35 $\,Myr while, for 
the 20\% of the sample with 
$t_{\rm sf} > 1 $\,Gyr, $\langle \Psi$($t_{\rm sf}$)$\rangle = 
25 h^{-2} \, M_{\odot}$~yr$^{-1}$\,.

This range of properties is further reflected in the total 
formed stellar masses $m_{\rm star}$ obtained by integrating
$\Psi$($t_{\rm sf}$) over $t_{\rm sf}$.
A variety of star formation histories was
considered (e.g. star formation which is continuous
or decreases with time); in general
$m_{\rm star}$ does not depend
sensitively on this choice, although
an older population of stars which by $z \simeq 3$
have faded at UV and optical wavelengths could remain hidden
(Papovich et al. 2001).
As can be seen from Figure 29, by redshift $z \simeq 3$ some galaxies
had apparently already assembled a stellar mass
comparable to that of an $L^{\ast}$ galaxy today,
$m_{\rm star}  \simeq 4 \times 10^{10} M_{\odot}$,
while 20\% of the sample have values of $m_{\rm star}$
one order of magnitude smaller.
These findings led Shapley et al. (2001)
to speculate that we may be beginning to discern an 
evolutionary sequence in Lyman break galaxies,
with the younger, dustier, more actively star-forming objects 
evolving to the older, less reddened, and more quiescent
phase. It remains to be seen how this scenario
stands up to the scrutiny of future observations,
as we try to link the properties of the stellar populations
of {\it individual} galaxies to other parameters, such as
dynamical mass and metallicity.

\subsection{Galactic-Scale Outflows}
%
%
\begin{figure}
\vspace*{-1.25cm}
\hspace*{-0.5cm}
\centering
\includegraphics[width=0.7\textwidth,angle=270]{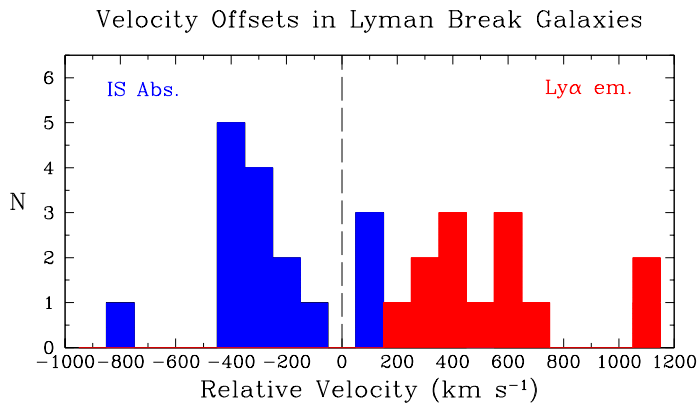}
\vspace*{-2.25cm}
\caption[]{
Velocity offsets of the interstellar
absorption lines (blue or dark grey)
and of the \lya\ emission line (red or light grey)
relative to [O~III] and \Hb.
Large scale motions of the order of several hundred
km~s$^{-1}$ are indicated by the systematic
tendency for the former to be blueshifted and the 
latter redshifted
relative to the nebular emission lines.
}
\label{}
\end{figure}

The near-IR spectroscopic survey by Pettini et al. (2001)
confirmed a trend which had already been 
suspected on the basis of the optical (rest-frame UV)
data alone.
When the redshifts of the interstellar absorption
lines, of the nebular emission lines, and of the 
resonantly scattered \lya\ emission line are compared within
the same galaxy, a systematic pattern
of velocity differences emerges in all LBGs
observed up to now (see Figure 30).
We interpret this effect as indicative of 
galaxy-wide outflows, presumably driven by
the supernova activity associated with the
star-formation episodes. 
Such `superwinds' appear to be
a common characteristic of galaxies with large rates of star 
formation per unit area at high, as well as low, redshifts
(e.g. Heckman 2001).
They involve comparable amounts of matter 
to that being turned into stars
(the mass outflow rate is of the same order as the star formation rate)
and about 10\% of the total kinetic energy delivered by the starburst
(Pettini et al. 2000b).
These outflows have a number of 
important astrophysical consequences.

First, they provide self-regulation to the star formation 
process---this is the `feedback'
required by theorists (e.g. Efstathiou 2000;
Binney, Gerhard, \& Silk 2001)
for realistic galaxy formation models.
Galactic 
winds may well be the key factor at 
the root of the `evolutionary
sequence' for LBGs just discussed (\S4.4).

Second, they can distribute the products of stellar
nucleosynthesis over large volumes of the intergalactic medium
since the outflow speeds are likely to exceed the escape 
velocities in many cases.
As we have seen, many LBGs are already metal-enriched 
at $z = 3$ and have by then been forming stars for much 
of the Hubble time. There is therefore at least
the potential for widespread pollution of 
the IGM with metals, thereby explaining
at least in part the results on the metallicity
of the \lya\ forest described in \S3.1 and 3.2).

Third, the outflowing hot gas is likely to `punch' through 
the neutral interstellar medium of the galaxies and provide
a route through which 
Lyman continuum photons can leak out of the galaxies,
easing the problem of how the universe came to be reionised
(Steidel, Pettini, \& Adelberger 2001).
Indeed it now appears (Adelberger et al. 2002, in preparation)
that LBGs have a substantial impact
on the surrounding IGM, and that shock-ionisation by their 
winds leads to a pronounced `proximity effect'---the
\lya\ forest is essentially cleared out by these outflows 
over radii of $\sim 100 h^{-1}$\,kpc.

\section{Bringing it All Together}

\subsection{A Global View of Metal Enrichment in the Universe\\
Two Billion Years after the Big Bang}

We could briefly summarise everything we have covered so far
as follows.
\begin{enumerate}
\item The intergalactic medium at $z = 3$ does not
consist entirely of pristine material. At least the
regions we have been able to probe so far
show traces of metals at levels between 1/1000 and 1/100
of solar metallicity. Matter in these regions, however,
is still at a higher density than the mean
density of the universe at that epoch, and it is unclear
at present whether a pre-galactic episode of star formation
(often referred to as Population III stars),
is required to explain the large scale distribution of 
elements in the IGM.

\item Damped \lya\ systems have abundances similar
to those of Population II stars in our Galaxy.
Perhaps they represent an early stage in the formation
of spiral galaxies, before most of the gas had been 
converted into stars. It is also clear that a wide
variety of galaxy morphologies, including low surface brightness
galaxies, share the common characteristic of providing
a large cross-section on the sky at high surface densities
of H~I.

\item Finally, Lyman break galaxies strongly resemble 
what we call Population I stars in the 
Milky Way. They are the sites of vigorous star formation 
which (i) has produced a relatively high level of 
chemical enrichment at early epochs,
(ii) has built up stellar masses of $\simgt 10^{10}\,M_{\odot}$
in a sizeable fraction of the population, 
and (iii) drives large-scale outflows of gas, metals and dust
into the intergalactic medium.
All of these characteristics point to LBGs as the 
progenitors of today's spiral bulges and elliptical
galaxies, observed during the most active phase 
in their lifetimes.
\end{enumerate}

%
%
\begin{figure}[]
\vspace*{0.5cm}
\centering
\resizebox{0.56\textwidth}{!}{\includegraphics{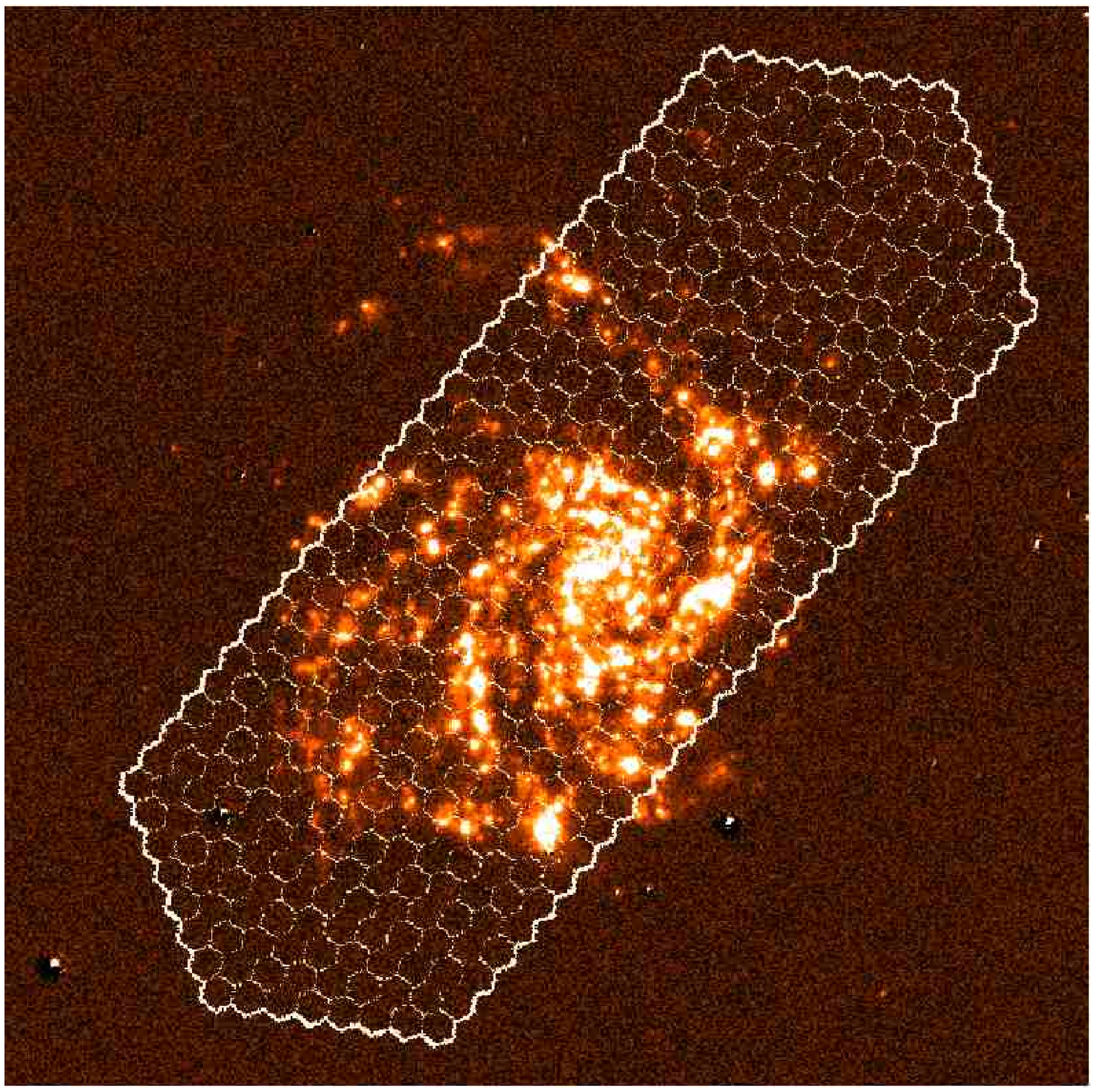}}
\resizebox{0.28\textwidth}{!}{\includegraphics{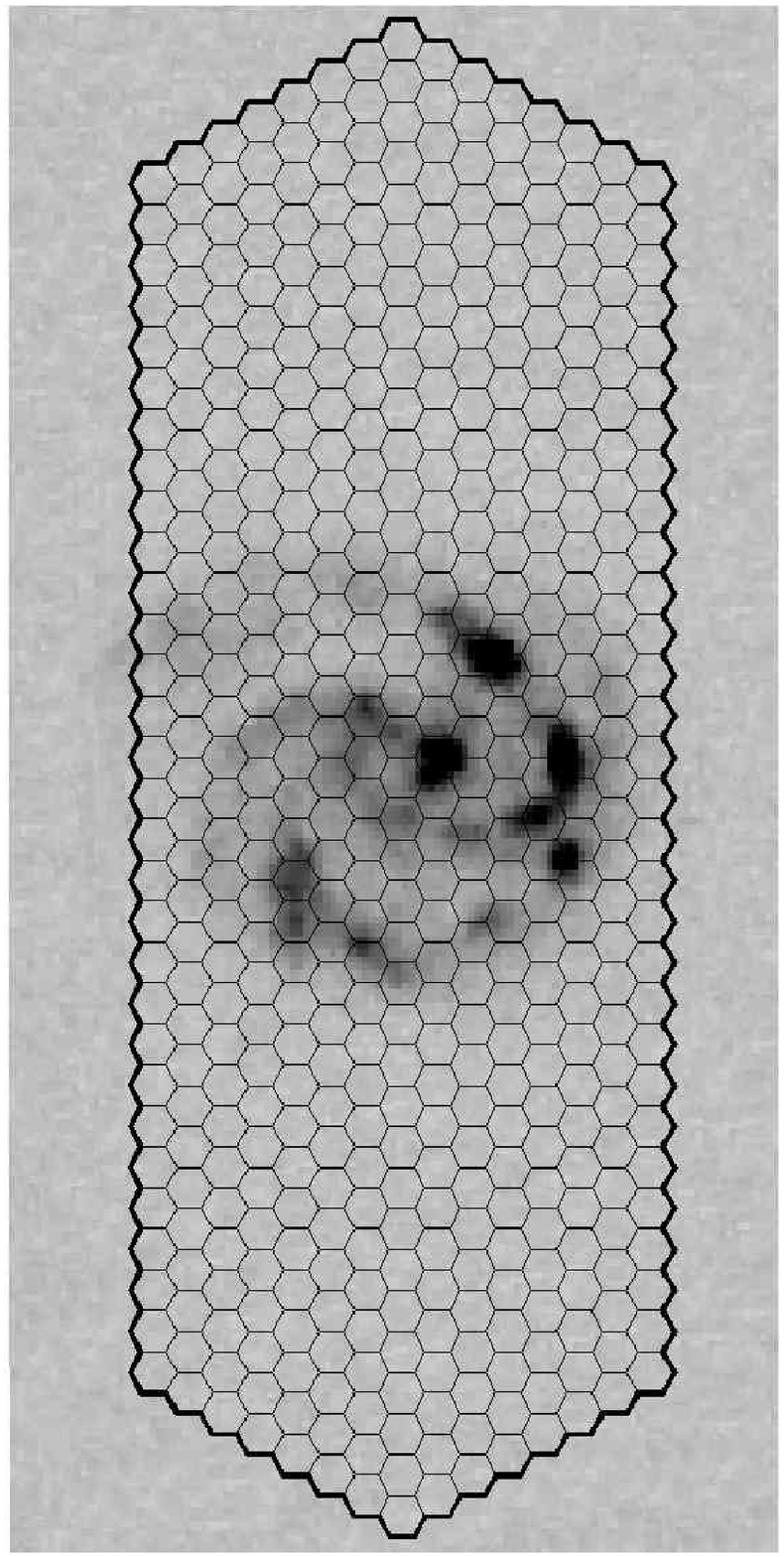}}
\vspace{0.5cm}
\caption{(Reproduced from Bunker et al. 2001).
{\it Left:\/} An H$\alpha$ image of the local spiral galaxy NGC\,4254 as
it would appear at $z=1.44$ with the CIRPASS integral field unit
overlaid (using 0.25\,arcsec diameter fibres). 
{\it Right:\/} A spiral galaxy at $z\approx 1$ from the Hubble Deep Field
$B$-band.  The star-forming H{\scriptsize~II} regions are prominent in
the rest-frame UV. CIRPASS will accurately determine the true star
formation rates, since (1) the compact knots of star formation are
well-matched to the fibre size, reducing the sky background and
increasing the sensitivity; (2) the large area surveyed by the integral
field unit covers most of a spiral disk and (3) the H$\alpha$ line is a
much more robust measure of the star formation rate than the
dust-suppressed UV continuum and resonantly-scattered Ly$\alpha$.
}
\end{figure}
%
%
%
\begin{center}
\vspace{0.75cm}
\begin{tabular}{l r r}
\multicolumn{3}{c}{{\bf Table 2.} Typical parameters of LBGs and DLAs at $z \simeq 3$}\\ 
\hline
\hline
\multicolumn{1}{l}{Property}&\multicolumn{1}{r}{~~~~~~~~~~~~LBGs}&\multicolumn{1}{r}{DLAs}\\
\hline
SFR ($M_{\odot}$~yr$^{-1}$)  & $\sim 50$  & $< 10$         \\
$Z$ ($Z_{\odot}$)            & $\sim 1/3$ & $\sim 1/20$    \\
$\Delta v$ (km~s$^{-1}$)     & $\sim 500$ & $\simlt 200$   \\
\hline
\end{tabular}
\end{center}
\vspace{0.5cm}
The connection between LBGs and DLAs is currently the subject of 
considerable discussion, as astronomers try and piece together
these different pieces of the puzzle describing the universe
at $z = 3$. Table 2 summarises some of the relevant properties.
Lyman break galaxies have systematically higher star formation rates and metallicities,
and their interstellar media have been stirred to higher velocities, than is the case
in most DLAs. These seemingly contrasting properties can perhaps be reconciled
if the two classes of objects are in fact drawn from the same luminosity 
function of galaxies at $z = 3$.
Since they are selected from magnitude limited samples,
the LBGs are preferentially bright galaxies---the data
in Table 2 refer to galaxies brighter than $L^{\ast}$
which corresponds to ${\cal R} = 24.5$ (Adelberger \& Steidel 2000).
If, on the other hand, the H~I absorption cross-section decreases
only slowly with galaxy luminosity, as is the case
at lower redshifts (Steidel, Dickinson, \& Persson 1994),
the DLA counts would naturally be dominated by the far more numerous
galaxies at the steep ($\alpha = -1.6$) faint end of the luminosity function.

Such a picture finds theoretical support in the results of hydrodynamical 
simulations and semi-analytic models of galaxy
formation (e.g. Nagamine et al. 2001; Mo, Mao, \& White 1999). 
In the coming years it will be tested by deeper and more extensive
searches for DLA galaxies, by comparing the clustering 
of LBGs and DLAs (Adelberger et al. 2002, in preparation),
and by more reliable measurements of the star formation activity
associated with DLAs. This last project is best tackled in the 
near-infrared, by targeting the H$\alpha$ emission line 
with integral field---rather than slit---spectroscopy,
as provided for example by the Cambridge Infrared
Panoramic Survey Spectrograph (CIRPASS---see Figure 31).

%
%
\begin{figure} 
\vspace*{-1.5cm} 
\hspace*{-1.0cm}
\psfig{figure=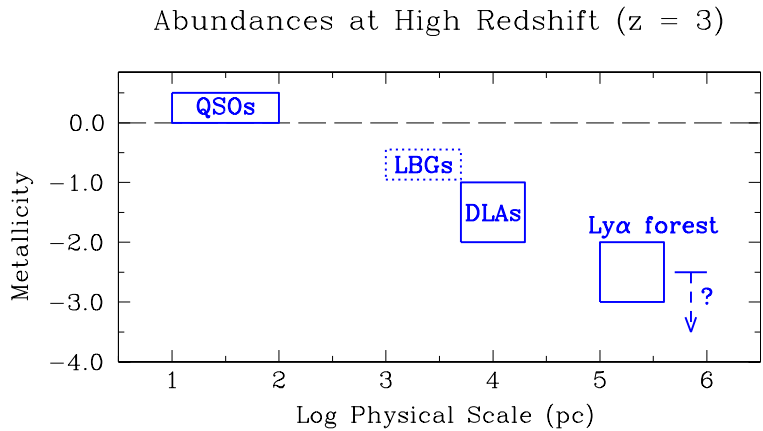,width=11cm,angle=270} 
\vspace*{-3.0cm} 
\caption[]{Summary of our current knowledge of
abundances at high redshift. The `metallicity' is plotted
on the $y$-axis on a log scale
relative to the solar reference; the latter is shown as 
the broken horizontal
line at 0.0 and corresponds to approximately
2\% of the baryons being incorporated in elements
heavier than helium.
The $x$-axis shows the typical linear dimensions
of the strucures to which the abundance 
measurements refer, from the central regions
of active galactic nuclei on scales of 10--100\,pc to
the intergalactic medium traced by the Ly\,$\alpha$ forest
on Mpc scales. Generally speaking, these typical linear scales are 
inversely proportional to the overdensities of the structures considered
relative to the background.
} 
\end{figure}

When we combine the available abundance determinations
for Lyman break galaxies, damped \lya\ systems and 
the \lya\ forest with those
for the inner regions of active galactic nuclei
(from analyses of the broad emission lines and  
outflowing gas in broad absorption line
QSOs---see Hamann \& Ferland 1999), a
`snapshot' of metal enrichment 
in the universe at $z \simeq 3$ 
emerges (Figure 32).
The $x$-axis in the figure gives the typical
linear dimensions of the structures to which
the abundance measurements refer, from the 10--100\,pc
broad emission line region of QSOs,
to the kpc scales of LBGs revealed by {\it HST} imaging
(Giavalisco, Steidel, \& Macchetto 1996), to the 10\,kpc
typical radii of DLAs deduced from their number density per unit redshift
(Steidel 1993), to the 100\,kpc dimensions of 
condensations in the \lya\ forest
with $N$(H~I)$\simgt 10^{14}$\,cm$^{-2}$ indicated
by the comparison of the absorption along adjacent sightlines
in the real universe (e.g. Bechtold et al. 1994) and in the
simulations (e.g. Hernquist et al. 1996).

These different physical scales in turn reflect the depths
of the underlying potential wells and therefore the overdensities
of matter in these structures relative to the mean density
of the universe. Even from such an approximate sketch as Figure 32,
it seems clear that it is this overdensity parameter
which determines the degree of metal enrichment
achieved at any particular cosmic epoch.
Thus, even at the relatively early times which correspond
to $z = 3$,  the gas 
in the deepest potential
wells where AGN are found had already undergone considerable
processing and reached solar or super-solar abundances. At the
other end of the scale, condensations in the \lya\ forest which
correspond to mild overdensities contained only traces of metals
with metallicity $Z \simeq 1/100 - 1/1000\,Z_{\odot}$. The
dependence of $Z$ on the environment appears to be stronger than
any age-metallicity relation---old does not necessarily mean
metal-poor, not only for our own Galaxy but also on a global
scale. This empirical conclusion can be understood in a general way
within the framework of hierarchical structure formation in cold
dark matter models (e.g. Cen \& Ostriker 1999; Nagamine et al.
2001).

\subsection{Missing Metals?}
%
%
\begin{figure} 
\vspace*{-1.5cm} 
\hspace*{-1.0cm}
\psfig{figure=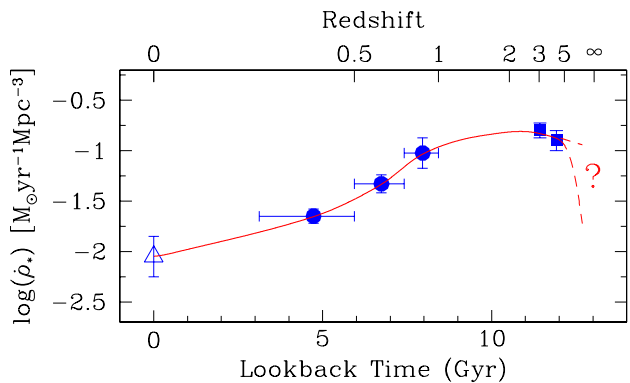,width=11cm,angle=270} 
\vspace*{-3.7cm} 
\caption[]{The comoving star formation rate density 
$\dot{\rho}_{\ast}$ vs.
lookback time compiled from wide angle, ground based
surveys (Steidel et al. 1999 and references therein).
The data shown here are for a $H_0 = 50$~km~s$^{-1}$~Mpc$^{-1}$,
$\Omega_{\rm M} = 1$, $\Omega_{\rm \Lambda} = 0$ cosmology.
}
\end{figure}

It can also be appreciated from Figure 32 that our knowledge
of element abundances at high redshift is still
very patchy.  This becomes all the more evident when we
attempt a simple counting exercise.
Figure 33 shows a recent version of a 
plot first constructed by Madau 
et al. (1996) which attempts to trace the `cosmic star formation 
history' by following the redshift evolution of the 
comoving luminosity density of star forming 
galaxies. 
This kind of plot enjoyed great popularity
after it was presented by Madau et al.;
more recently astronomers have approached it
with greater caution as they have become 
more aware of the 
uncertainties involved. In particular,
the dust corrections to the 
data in Figure 33
have been the subject of intense debate
over the last five years, as has been the 
contribution to $\dot{\rho}_{\ast}$ 
from galaxies which may be obscured at 
visible and ultraviolet wavelengths
and only detectable in the sub-mm regime
with instruments such as SCUBA.
Furthermore, the normalisation of
the plot depends on the IMF
and on the slope of the faint end of the 
galaxy luminosity function.
Nevertheless, if we assume 
that we have the story about right, 
some interesting consequences follow.

The first question one may ask is: {\it ``What is the total mass in stars
obtained by integrating under the curve in Figure 33?''}
The data points in Figure 33 were derived assuming a Salpeter IMF
(with slope $-2.35$) between $M = 100~M_{\odot}$ and $0.1~M_{\odot}$\,.
However, for a more realistic IMF which flattens 
below $1~M_{\odot}$, the values of $\dot{\rho}_{\ast}$
in Figure 33 should be reduced by a factor of 2.5\,.\footnote{I
have not applied this correction directly to 
Figure 33 in order to ease the comparison with earlier versions of 
this plot.} With this correction:

\begin{equation}
	\int_{0}^{13~Gyr}\dot{\rho}_{\ast}\hspace{-0.8mm}'~dt \simeq
	3.3 \times 10^8~M_{\odot}~{\rm Mpc}^{-3} =
	0.0043~\rho_{\rm crit} \approx \Omega_{\rm stars}~\rho_{\rm crit}
	\label{}
\end{equation}

\noindent where 
$\rho_{\rm crit} = 3.1 \times 10^{11}\,h^2\,M_{\odot}\,{\rm Mpc}^{-3}$
and $\Omega_{\rm stars} \simeq {\rm (}1.5 - 3{\rm )}\,h^{-1} \times 10^{-3}$  
(Cole et al. 2001) is the fraction of the closure density
contributed by stars at $z = 0$.
Thus, within the rough accuracy with which this accounting can 
be done, the star formation history depicted in Figure 33 is apparently 
sufficient to produce the entire stellar 
content, in disks and spheroids, of the present day universe.
Note also that $\approx 1/4$ of today's stars were
made before $z = 2.5$ (the uncertain 
extrapolation of $\dot{\rho}_{\ast}$ beyond $z = 4$ 
makes little difference).

We can also ask: {\it ``What is the total mass of metals produced by $z = 
2.5$?''}.
Using the conversion factor 
$\dot{\rho}_{\rm metals} = 1/42~\dot{\rho}_{\ast}$
to relate the comoving density of synthesised metals to
the star formation rate density (Madau et al. 1996)
we find 

\begin{equation}
	\int_{11~Gyr}^{13~Gyr}\dot{\rho}_{\rm metals}~dt \simeq
	4.5 \times 10^6~M_{\odot}~{\rm Mpc}^{-3} 
\end{equation}
which corresponds to
\begin{equation}
        \Omega_Z \simeq 0.035 \times {\rm (}\Omega_{\rm baryons} \times 0.0189 {\rm )}
	\label{}
\end{equation}
\noindent where  
$\Omega_{\rm baryons} = 0.088$ for $h = 0.5$ (\S1.1)
and 0.0189 is the mass fraction of elements heavier than helium
for solar metallicity (Grevesse \& Sauval 1998).
In other words, the amount of metals produced by the star formation we 
{\it see} at high redshift (albeit corrected for dust extinction)
is sufficient to enrich the whole baryonic content of the universe 
at $z = 2.5$ to $\approx 1/30$ of solar metallicity.
Note that this conclusion does not depend sensitively on the IMF.

%
%
\begin{figure}
\vspace*{-7.35cm}
\hspace*{-2.75cm}
\psfig{figure=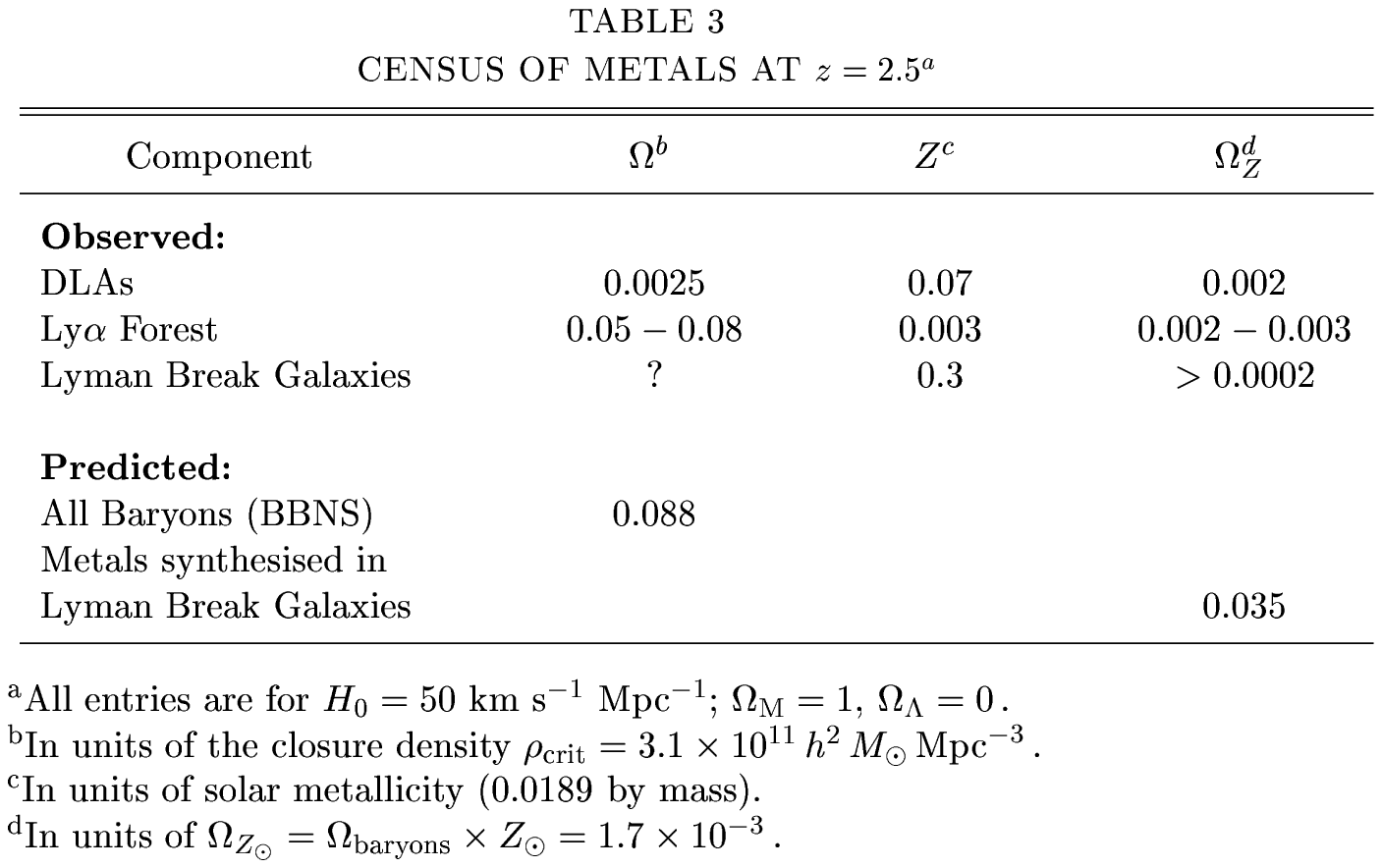,width=17.5cm,angle=0}
\vspace{-9.75cm}
\end{figure}

As can be seen from Table 3, this leaves us with a serious 
`missing metals' problem which has also been discussed 
in more detail by Pagel (2002).
The metallicity of damped \lya\ systems is in the right ballpark,
but $\Omega_{\rm DLA}$ is only a small fraction of $\Omega_{\rm baryons}$. 
Conversely, while the \lya\ forest may account for a large fraction of the baryons, 
its metal content is one order of magnitude too low.
The contribution of Lyman break galaxies to the cosmic inventory
of metals is even more uncertain. The value in Table 3 is
a strict (and not very informative) lower limit, calculated
from the luminosity function of Steidel et al. (1999), taking
into account {\it only} galaxies brighter than $L^{\ast}$
and assigning to each a mass
$M^{\ast} = 10^{11}\,M_{\odot}$ (which is likely to 
be a lower limit, as discussed by Pettini et al. 2001)
and metallicity $Z = 1/3\,Z_{\odot}$. 
Galaxies fainter than $L^{\ast}$ are not included
in this census because we still have no idea of their 
metallicities; potentially they could make a significant
contribution to $\Omega_Z$(LBG) because they are so numerous.

Nevertheless, when we add up all the metals which have been measured
with some degree of confidence up to now, we find that they account
for no more than $\approx 10-15$\% of what we expect to have been produced 
by $z = 2.5$ (last column of Table 3). Where are these missing metals?
Possibly, $\Omega_Z$(DLA) has been underestimated, if the dust
associated with the most metal-rich DLAs obscures background QSOs
sufficiently to make them drop out of current samples.
However, preliminary indications based on the CORALS survey 
by Ellison et al. (2001) suggest that this may be a relatively
minor effect (see also Prochaska \& Wolfe 2002).
The concordance in the values of $\Omega_Z$(IGM) derived from observations
of O~VI and C~IV absorption in the \lya\ forest makes it unlikely
that the metallicity of the widespread IGM has been underestimated
by a large factor. On the other hand, we do know that Lyman
break galaxies commonly drive large scale outflows;
it is therefore possible, and indeed likely, that they 
enrich with metals much larger masses of gas than those 
seen directly as sites of star formation.
This gas and associated metals may be difficult to detect
if they are at high temperatures, and yet may make a
major contribution to $\Omega_Z$; there are tantalising hints
that this could be the case 
at the present epoch (Tripp, Savage, \& Jenkins 2000;
Mathur, Weinberg, \& Chen 2002).

In concluding this series of lectures, it is clear that
while we have made some strides forward towards
our goal of charting the chemical history of the
universe, our task is far from complete. It is my hope
that, stimulated in part by this school,
some of the students who have attended it
will soon be contributing
to this exciting area of observational
cosmology as their enter their research careers.\\

I am very grateful to C\'{e}sar Esteban, Artemio Herrero, Rafael Garc\'{\i}a L\'{o}pez and
Prof. Francisco S\'{a}nchez for inviting me to take part in a very enjoyable Winter
School, and to the students for their patience and challenging questions.
The results described in these lectures were obtained in various collaborative 
projects primarily with Chuck Steidel, Kurt Adelberger, David Bowen, 
Mark Dickinson, Sara Ellison, Mauro Giavalisco, Samantha Rix and
Alice Shapley; I am fortunate indeed to be working with such 
productive and generous colleagues. 
Special thanks to Alec Boksenberg and Bernard Pagel for continuing inspiration and
for valuable comments on an early version of the manuscript.
As can be appreciated from these
lecture notes, the measurement of element abundances at high redshifts
is a vigorous area of research. In the spirit of the school,
I have not attempted to give a comprehensive set of references
to all the numerous papers on the subject which have appeared
in recent years, as one would in a review. 
Rather, I have concentrated on the main issues and
only given references as pointers for further reading. I apologise
for the many excellent papers which have therefore been omitted
from the (already long) list of references. Such omissions do not 
in any way denote criticism on my part of the work in question.


\begin{thebibliography}{}

\bibitem[]{}{\sc Adelberger, K.L., \& Steidel, C.C.} 2000, ApJ, 544, 218

\bibitem[]{}{\sc Aguirre, A.,  Hernquist, L.,  Schaye, J.,
Katz, N.,  Weinberg, D. H., \& Gardner, J.} 2001, ApJ, 561, 521

\bibitem[]{}{\sc Bechtold, J., Crotts, A.P.S., Duncan, R.C., \&
Fang, Y.} 1994, ApJ, 437, L83

\bibitem[]{}{\sc Bechtold, J.} 2002, in Galaxies at High Redshift, 
eds. I. P\'erez-Fournon, M. Balcells, F. Moreno-Insertis, \&
F. S\'anchez, (Cambridge: Cambridge Univ. Press), p. 131

\bibitem[]{}{\sc  Binney, J., Gerhard, O., \& Silk, J.} 2001,
MNRAS, 321, 471

\bibitem[]{}{\sc Boissier, S., P\'{e}roux, C., \& Pettini, M.} 2002,
MNRAS, submitted

\bibitem[]{} {\sc Bouch\'{e}, N., Lowenthal, J.D., Charlton, J.C., Bershady, M.A.,
Churchill, C.W., \& Steidel, C.C.} 2001, ApJ, 550, 585

\bibitem[]{} {\sc Bouchet, P., Lequeux, J., Maurice, E., Prevot, L., \& 
Prevot-Burnichon, M.L.} 1985, A\&A, 149, 330

\bibitem[]{} {\sc Bruzual, A.G., \& Charlot, S.} 1993, ApJ, 405, 538

\bibitem[]{} {\sc Bunker, B., Ferguson, A., Johnson, R. et al.} 2001, in Deep Fields,
ESO Astrophysics Symposia, eds. S. Cristiani, A. Renzini, \& R.E. Williams,
(Berlin:Springer), 330

\bibitem[]{} {\sc Carswell, R.F., Schaye, J., \& Kim, T.S.} 2002, ApJ, in press (astro-ph/0204370)

\bibitem[]{}{\sc Cen, R., \& Ostriker, J.P.} 1999, ApJ, 519, L109

\bibitem[]{}{\sc Cole, S., Norberg, P., Baugh, C.M., et al.} 2001, MNRAS, 326, 255

\bibitem[]{}{\sc  Cowie, L.L., Songaila, A., Kim, T.S., \& Hu, E.M.}
1995, AJ, 109, 1522

\bibitem[]{} {\sc Croft, R.A.C., Weinberg, D.H., Bolte, M. et al.} 2002,
ApJ, submitted (astro-ph/0012324)

\bibitem[]{} {\sc Dav\'{e}, R., Hellsten, U., Hernquist, L., 
Katz, N. \& Weinberg, D.H.} 1998, ApJ, 509, 661 

\bibitem[]{}{\sc Diplas, A., \& Savage, B.D.} 1994, ApJ, 427, 274

\bibitem[]{}{\sc Edmunds, M.G., \& Pagel, B.E.J.} 1978, MNRAS, 185, 77P


\bibitem[]{}{\sc Efstathiou, G.} 2000, MNRAS, 317, 697

\bibitem[]{}{\sc  Ellingson, E., Yee, H.K.C., Bechtold, J., \& Elston, R.} 1996,
ApJ, 466, L71

\bibitem[]{}{\sc Ellison, S.L.} 2000, Ph.D. Thesis, University of Cambridge

\bibitem[]{}{\sc  Ellison, S.L., Lewis, G.F., Pettini, M., Chaffee, F.H., 
\& Irwin, M.J.} 1999, ApJ, 520, 456

\bibitem[]{} {\sc Ellison, S.L., Songaila, A., Schaye, J., Cowie, L.L., 
\& Pettini, M.} 2000, AJ, 120, 1175

\bibitem[]{}{\sc Ellison, S.L., Yan, L., Hook, I.M., 
Pettini, M., Wall, J.V., \& Shaver, P.} 2001,
A\&A, 379, 393

\bibitem[]{}{\sc Fall, S.M.} 1996, in The Hubble Space Telescope and the 
High Redshift Universe, eds. N. Tanvir, A. Aragon-Salamanca, \& J.V. Wall
(Singapore: World Scientific), 303

\bibitem[]{}{\sc Fuhrmann, K.} 1998, A\&A, 338, 161


\bibitem[]{}{\sc Garnett, D.R., Shields, G.A., Peimbert, M., et al.} 1999, ApJ, 513, 168

\bibitem[]{}{\sc Giavalisco, M., Steidel, C.C., \& Macchetto, F.D.} 1996, ApJ, 470, 189

\bibitem[]{}{\sc Gilmore, G., \& Wyse, R.F.G.} 1991, ApJ, 367, L55

\bibitem[]{}{\sc Grevesse, N., \& Sauval, A.J.} 1998, Space Sci Rev, 85, 161

\bibitem[]{}{\sc Hamann, F., \& Ferland, G.} 1999, ARA\&A, 37, 487

\bibitem[]{}{\sc Haehnelt, M.G., Steinmetz, M., 
\& Rauch, M.} 1998, ApJ, 495, 647

\bibitem[]{}{\sc Heckman, T.M.} 2001, in ASP Conf. Ser.,
Gas and Galaxy Evolution, ASP Conference Proceedings, Vol. 240,
eds. J.E. Hibbard, M.P. Rupen, \& J.H. van Gorkom,
(San Francisco:ASP), 345

\bibitem[]{}{\sc Henry, R.B.C., Edmunds, M.G., \& K\"{o}ppen, J.} 2000, ApJ, 541, 660

\bibitem[]{}{\sc Hernquist, L., Katz, N., Weinberg, D.H., \&
Miralda-Escud\'{e}, J.} 1996, ApJ, 457, L51

\bibitem[]{}{\sc Holweger, H.} 2001, in Solar and Galactic Composition,
ed. R.F. Wimmer-Schweingruber,  American Institute of Physics Conference
proceedings, 598, 23

\bibitem[]{}{\sc Kanekar, N., \&  Chengalur, J.N.} 2001, A\&A, 369, 42

\bibitem[]{}{\sc Kennicutt, R.C.} 1998a, in ASP Conf. Ser. 142,
The Stellar Initial Mass Function,
ed. G. Gilmore \& D. Howell (San Francisco: ASP), 1

\bibitem[]{}{\sc Kennicutt, R.C.} 1998b, in 
The Next Generation Space Telescope: Science Drivers and Technological Challenges, 
34th Li\`{e}ge Astrophysics Colloquium, 81

\bibitem[]{}{\sc Krauss, L.M., \& Chaboyer, B.} 2001, Nature, submitted (astro-ph/0111597)

\bibitem[]{}{\sc Kulkarni, V., \& Fall, S.M.}, 2002, ApJ, submitted 

\bibitem[]{}{\sc Laird, J.B., Rupen, M.P., Carney, B.W., \& Latham, D.W.}
1988, AJ, 96, 1908

\bibitem[]{}{\sc Lanzetta, K.M.} 1993, in 
The Environment and Evolution of Galaxies, 
eds. J.M. Shull \& H.A. Thronson,
(Dordrecht: Kluwer), 237

\bibitem[]{}{\sc Larsen, T.I., Sommer-Larsen, J., \& Pagel, B.E.J.} 2001, MNRAS, 323, 555

\bibitem[]{}{\sc Ledoux, C., Bergeron, J., \& Petitjean, P.} 2002, A\&A, 385, 802

\bibitem[]{}{\sc Ledoux, C., Petitjean, P., Bergeron, J.,  
Wampler, E.J., \& Srianand, R.} 1998, A\&A, 337, 51

\bibitem[]{}{\sc Leitherer, C., Schaerer, D., Goldader, J.D., et al.} 1999, ApJS, 123, 3

\bibitem[]{}{\sc Leitherer, C.,  Le\~{a}o, J.R.S., Heckman, T.M., Lennon, D.J., 
Pettini, M., \&  Robert, C.} 2001, ApJ, 550, 724

\bibitem[]{}{\sc Madau, P., Ferguson, H.C., Dickinson, M.E., Giavalisco, 
M., Steidel, C.C., \& Fruchter, A.} 1996, MNRAS, 283, 1388

\bibitem[]{}{\sc Madau, P., Ferrara, A., \& Rees, M.J.} 2001, ApJ, 555, 92

\bibitem[]{}{\sc Madau, P., \& Shull, J.M.} 1996, ApJ, 457, 551 

\bibitem[]{}{\sc Mathur, S., Weinberg, D.H., \& Chen, X.} 2002, ApJ, submitted (astro-ph/0206121)

\bibitem[]{}{\sc Matteucci, F., \& Recchi, S.} 2001, ApJ, 558, 351

\bibitem[]{}{\sc McGaugh, S.} 1991, ApJ, 380, 140

\bibitem[]{}{\sc McWilliam, A., Preston, G.W., Sneden, C., \& Searle, L.} 1995, AJ, 109, 2757

\bibitem[]{}{\sc Meyer, D.M., Welty, D.E., \& York, D.G.} 1989, ApJ, 343, L37

\bibitem[]{}{\sc Meynet, G. \& Maeder, A.} 2002, A\&A, 381, L25

\bibitem[]{}{\sc Miralda-Escud\'{e}, J., \& Rees, M.J.} 1997, ApJ, 478, L57

\bibitem[]{}{\sc Mo, H.J., Mao, S., \& White, S.D.M.} 1999, MNRAS, 304, 175

\bibitem[]{}{\sc Molaro, P., Bonifacio, P.,  Centuri\'{o}n, M.,
D'Odorico, S., Vladilo, G., Santin, P., \& Di Marcantonio, P.} 2000, ApJ, 541, 54

\bibitem[]{}{\sc Molaro, P., Levshakov, S.A.,  D'Odorico, S.,
Bonifacio, P., \&  Centuri\'{o}n, M.} 2001, ApJ, 549, 90

\bibitem[]{}{\sc M\o ller, P., Warren, S.J., Fall, S.M., Fynbo, J.U., 
\& Jakobsen, P.} 2002, ApJ, in press (astro-ph/0203361)

\bibitem[]{}{\sc  Nagamine, K., Fukugita, M., Cen, R., \& Ostriker, J.P.} 2001,
ApJ, 558, 497

\bibitem[]{}{\sc  Pagel, B.E.J.} 2002, in ASP Conf. Series 253,
Chemical Enrichment of Intracluster and Intergalactic Medium,
eds. R. Fusco-Femiano \& F. Matteucci, (San Francisco: ASP), 489

\bibitem[]{}{\sc  Pagel, B.E.J., Edmunds, M.G., Blackwell, D.E.,
Chun, M.S., \& Smith, G.} 1979, MNRAS, 189, 95

\bibitem[]{}{\sc  Pagel, B.E.J., \& Tautvaisiene, G.} 1998, MNRAS, 299, 535

\bibitem[]{}{\sc  Papovich, C., Dickinson, M., \& Ferguson, H.C.} 2001,
ApJ, 559, 620

\bibitem[]{}{\sc Pei, Y.C., Fall, S.M., \& Bechtold, J.} 1991, ApJ, 378, 6

\bibitem[]{}{\sc Pei, Y.C., Fall, S.M., \& Hauser, M.G.} 1999, ApJ, 522, 604

\bibitem[]{}{\sc Penton, S.V., Shull, J.M., \& Stocke, J.T.} 2000, ApJ, 544, 150

\bibitem[]{}{\sc P\'{e}roux, C., McMahon, R.G., Storrie-Lombardi, L.J., 
\& Irwin, M.J.} 2002, MNRAS, submitted (astro-ph/0107045)

\bibitem[]{}{\sc Petitjean, P., \& Bergeron, J.} 1994,  A\&A, 283, 759

\bibitem[]{}{\sc Petitjean, P., Srianand, R., \& Ledoux, C.} 2000, A\&A, 364, L26

\bibitem[]{}{\sc Pettini, M.} 2001, in
Gaseous Matter in Galaxies and Intergalactic Space,
ed. R. Ferlet, M. Lemoine, J.M. Desert, \& B. Raban
(Frontier Group), 315

\bibitem[]{}{\sc Pettini, M., Boksenberg, A., \& Hunstead, R.W.} 1990, ApJ, 348, 48

\bibitem[]{}{\sc Pettini, M., \& Bowen, D.V.} 2001, ApJ, 560, 41

\bibitem[]{}{\sc Pettini, M., Ellison, S. L., Bergeron, J., \& Petitjean, P.} 2002a, A\&A, in press (astro-ph/0205472)

\bibitem[]{}{\sc Pettini, M., Ellison, S. L., Steidel, C. C., \& Bowen, D. V.} 1999, ApJ, 510, 576

\bibitem[]{}{\sc Pettini, M., Ellison, S. L., Steidel, C. C.,
Shapley, A. E., \& Bowen, D. V.} 2000a, ApJ, 532, 65 

\bibitem[]{}{\sc Pettini, M., Lipman, K., \& Hunstead, R.W.} 1995, ApJ, 451, 100

\bibitem[]{}{\sc Pettini, M., King, D.L., Smith, L.J., 
\& Hunstead, R.W.} 1997a, ApJ, 478, 536

\bibitem[]{}{\sc Pettini, M., Rix, S.A., Steidel, C.C., Adelberger, K.L., Hunt, M.P.,
\& Shapley, A.E.} 2002b, ApJ, 569, 742

\bibitem[]{}{\sc Pettini, M., Shapley, A.E., Steidel, C.C., Cuby, J.G., 
Dickinson, M., Moorwood, A.F.M., Adelberger, K.L., \& Giavalisco, M.}
2001, ApJ, 554, 981

\bibitem[]{}{\sc Pettini, M., Smith, L.J., Hunstead, R.W., \& King, D.L.} 1994, ApJ, 426, 79

\bibitem[]{}{\sc Pettini, M., Smith, L.J., King, D.L. \& Hunstead, R.W.} 1997b, ApJ, 486, 665

\bibitem[]{}{\sc  Pettini, M., Steidel, C.C., Adelberger, K.L., Dickinson, M., 
\& Giavalisco, M.} 2000b, ApJ, 528, 96

\bibitem[]{}{\sc Prochaska, J.X., Gawiser, E., \& Wolfe A.} 2001, ApJ, 552, 99

\bibitem[]{}{\sc Prochaska, J.X., Gawiser, E., \& Wolfe A., et al.} 2002, AJ, 123, 2206

\bibitem[]{}{\sc Prochaska, J.X., \& Wolfe A.} 1998, ApJ, 507, 113

\bibitem[]{}{\sc Prochaska, J.X., \& Wolfe A.} 1999, ApJS, 121, 369

\bibitem[]{}{\sc Prochaska, J.X., \& Wolfe A.} 2002, ApJ, 566, 68

\bibitem[]{}{\sc  Rao, S.M., \& Turnshek, D.A.} 2000, ApJS, 130, 1

\bibitem[]{}{\sc  Rauch, M.} 1998, ARA\&A, 36, 267

\bibitem[]{}{\sc Rosenberg, J., \& Schneider, S.} 2002, 
ApJ, submitted (astro-ph/0202216)

\bibitem[]{}{\sc Ryan, S.G., Norris, J.E., \& Beers, T.C.} 1996, ApJ, 471, 254

\bibitem[]{}{\sc Sargent, W.L.W., Young, P.J., Boksenberg, A., \& Tytler, D.} 1980,
ApJS, 42, 41

\bibitem[]{}{\sc Savage, B.D., \& Sembach, K.R.} 1996, ARA\&A, 34, 279

\bibitem[]{}{\sc  Schaye, J., Rauch, M., Sargent, W.L.W., \& Kim, T.S.}
2000, ApJ, 541, L1

\bibitem[]{}{\sc Seitz, S., Saglia, R.P., Bender, R., Hopp, U., Belloni, P., \&
Ziegler, B. } 1998, MNRAS, 298, 945

\bibitem[]{}{\sc Shapley, A.E., Steidel, C.C., Adelberger, K.L.,
Dickinson, M., Giavalisco, M., \& Pettini, M.} 2001, ApJ, 562, 95

\bibitem[]{}{\sc Sneden, C., Gratton, R.G., \& Crocker, D.A.} 1991, A\&A, 246, 354

\bibitem[]{}{\sc Songaila, A.} 1997, ApJ, 490, L1

\bibitem[]{}{\sc Songaila, A.} 2001, ApJ, 561, L153

\bibitem[]{}{\sc Songaila, A., \& Cowie, L.L.} 2002, AJ, 123, 2183

\bibitem[]{}{\sc Steidel, C.C.} 1993, in 
The Environment and Evolution of Galaxies, 
eds. J.M. Shull \& H.A. Thronson,
(Dordrecht: Kluwer), 263

\bibitem[]{}{\sc Steidel, C.C.} 2000, in Discoveries and Research Prospects 
from 8- to 10-Meter-Class Telescopes, ed. J. Bergeron, Proc. SPIE Vol. 4005, 22

\bibitem[]{} {\sc Steidel, C.C., Adelberger, K.L., 
Giavalisco, M., Dickinson, M., \& Pettini, M.} 1999, ApJ, 519, 1

\bibitem[]{}{\sc Steidel, C.C., Dickinson, M., Meyer, D.M.,
Adelberger, K. L., \& Sembach, K.R.} 1997, ApJ, 480, 568

\bibitem[]{}{\sc Steidel, C.C., Dickinson, M., \& Persson, S.E.} 1994, ApJ, 437, L75

\bibitem[]{}{\sc Steidel, C.C., Giavalisco, M., Pettini, M., Dickinson, M.,
\& Adelberger, K. L.} 1996, ApJ, 462, L17

\bibitem[]{}{\sc Steidel, C.C., Pettini, M., \& Adelberger, K.L.} 2001, 
ApJ, 546, 665

\bibitem[]{}{\sc Steidel, C.C., Pettini, M., \& Hamilton, D.} 1995, AJ, 110, 2519

\bibitem[]{}{\sc Storrie-Lombardi, L.J., \& Wolfe, A.M.} 2000, ApJ, 543, 552

\bibitem[]{}{\sc Teplitz, H.I., McLean, I.S., Becklin, E.E., et al.} 2000, ApJ, 533, L65

\bibitem[]{}{\sc Tripp, T.M., Savage, B.D., \& Jenkins, E.B.} 2000, ApJ, 534, L1

\bibitem[]{}{\sc Turner, M.S.} 2002, in Proceedings of the XXth International Symposium on Photon and 
Lepton Interactions, in press (astro-ph/0202007)

\bibitem[]{}{\sc Tytler, D.} 1987, ApJ, 321, 49

\bibitem[]{}{\sc  Tytler, D., Fan, X.-M., Burles, S., Cottrell, L., Davis, C.,
Kirkman, D., \& Zuo, L.} 1995, QSO Absorption Lines, 
ed. G. Meylan,  (Garching, ESO), 289

\bibitem[]{}{\sc Tytler, D., O'Meara, J.M., Suzuki, N., \& Lubin, D.} 2000,
Physica Scripta T, 85, 12

\bibitem[]{}{\sc Vladilo, G.} 2002a, ApJ, 569, 295

\bibitem[]{}{\sc Vladilo, G.} 2002b, A\&A, in press (astro-ph/0206048)

\bibitem[]{}{\sc Vladilo, G., Centuri\'{o}n, M., Bonifacio, P., \& Howk, J.C.} 2001, ApJ, 557, 1007

\bibitem[]{}{\sc Weinberg, D.H., Katz, N., \& Hernquist, L.} 1998, 
in ASP Conf. Series 128, Origins, ed. C.E. Woodward, J.M. Shull, 
\& H.A. Thronson, (San Francisco: ASP), 21

\bibitem[]{}{\sc Wheeler, J.C., Sneden, C., \& Truran, J.W.} 1989, ARA\&A, 27, 279

\bibitem[]{}{\sc Wolfe, A. M., Turnshek, D. A.,  Smith, H. E., \& Cohen, R. D.} 1986, ApJS, 61, 249

\bibitem[]{}{\sc Wolfire, M.G., Hollenbach, D., McKee, C.F., Tielens, A.G.G.M., \& Bakes, E.L.O.} 1995, ApJ, 443, 152

\bibitem[]{}{\sc Wyse, R.F.G., \& Gilmore, G.} 1995, AJ, 110, 2771

\bibitem[]{}{\sc Yee, H.K.C., Ellingson, E., Bechtold, J., Carlberg, R.G., 
\& Cuillandre, J.-C.} 1996, AJ, 111, 1783


\end{thebibliography}
\end{document}